\documentclass[amsmath,amssymb]{revtex4}
\usepackage{graphicx}

\usepackage{bm}

\begin{document}

\newcommand{\uu}[1]{\underline{#1}}
\newcommand{\pp}[1]{\phantom{#1}}
\newcommand{\be}{\begin{eqnarray}}
\newcommand{\ee}{\end{eqnarray}}
\newcommand{\ve}{\varepsilon}
\newcommand{\vs}{\varsigma}
\newcommand{\Tr}{{\,\rm Tr\,}}
\newcommand{\pol}{\textstyle{\frac{1}{2}}}
\newcommand{\lbar}{l_{^{\!\bar{}}}}

\title{Automatic regularization by quantization in reducible representations of CCR: Point-form quantum optics with classical sources}
\author{Marek Czachor$^{1,2}$ and Klaudia Wrzask$^1$}
\affiliation{
$^1$ Katedra Fizyki Teoretycznej i Informatyki Kwantowej\\
Politechnika Gda\'nska, 80-952 Gda\'nsk, Poland\\
$^2$ Centrum Leo Apostel (CLEA)\\
Vrije Universiteit Brussel, 1050 Brussels, Belgium
}

\begin{abstract}
Electromagnetic fields are quantized in a manifestly covariant way by means of a class of reducible ``center-of-mass $N$-representations" of the algebra of canonical commutation relations (CCR). The four-potential $A_a(x)$ transforms in these representations as a Hermitian four-vector field in Minkowski four-position space (without change of gauge), but in momentum space it splits into spin-1 massless photons and two massless scalars. What we call quantum optics is the spin-1 sector of the theory. The scalar fields have physical status similar to that of dark matter (spin-1 and spin-0 particle numbers are separately conserved). There are no negative-norm or zero-norm states. Unitary dynamics is given by the point-form interaction picture, with minimal-coupling Hamiltonian constructed from fields that are free on the null-cone boundary of the Milne universe. SL(2,C) transformations as well as the dynamics are represented unitarily in the Hilbert space corresponding to $N$ four-dimensional oscillators. Vacuum is a Bose-Einstein condensate of the $N$-oscillator gas and is given by any $N$-oscillator product state annihilated by all annihilation operators. The form of $A_a(x)$ is determined by an analogue of the twistor equation. The same equation guarantees that the set of vacuum states is Poincar\'e invariant. The formalism is tested on quantum fields produced by pointlike classical sources. Photon statistics is well defined even for pointlike charges, with ultraviolet and infrared regularizations occurring automatically as a consequence of the formalism. The probabilities are not Poissonian but of a R\'enyi type with $\alpha=1-1/N$; the Shannon limit $N\to\infty$ is an ultraviolet/infrared-regularized Poisson distribution. The average number of photons occurring in Bremsstrahlung splits into two parts: The one due to acceleration, and the one that remains nonvanishing even for inertially moving charges. Classical Maxwell electrodynamics is reconstructed from coherent-state averaged solutions of Heisenberg equations. We show in particular that static pointlike charges polarize vacuum and produce effective charge densities and fields whose form is sensitive to both the choice of representation of CCR and the corresponding vacuum state.
\end{abstract}
\maketitle

\section{Introduction}

What we present below is a manifestly covariant quantum theory of light with positive-definite metric in a Hilbert space and Hermitian four-potential, where regularizations occur automatically and do not have to be motivated by hand-waving arguments. We abandon the ``Zeroth Axiom" of quantum field theory stating that there is only one relativistically invariant vacuum state, and show how to construct a theory where the vacuum is represented by a relativistically invariant Hilbert subspace of states annihilated by all annihilation operators.

The dynamics is not given by Maxwell's equations, but by the Heisenberg equation. The formalism is gauge independent in the following sense: The requirement that the set of states annihilated by all annihilation operators is SL(2,C) invariant almost uniquely (i.e. up to a solution of a twistor-like equation) fixes the form of the four potential. All potentials within this class produce the same physical results not only for the $S$-matrix, but also for finite-time evolutions. The dynamics is unitary and well defined even for pointlike classical sources, and reconstructs Maxwell electrodynamics at the level of coherent-state averages.
The time $0\leq \tau$ that parametrizes the evolution is counted since the origin of the Universe.

We have tried to follow the philosophy of P. A. M. Dirac expressed in his talk given on 1 July 1982 at the Lindau gathering of Nobel laureates. Criticizing the general paradigm of quantum field theory he said:

``I feel that we have to insist on the validity of this Heisenberg equation.
This is the whole basis of quantum theory. We have got to hold onto it
whatever we do, and if the equation gives results which are not correct
it means that we are using the wrong Hamiltonian. This is the point I
want to emphasise (...). Heisenberg originally formulated these equations
with the dynamical variables appearing as matrices. You can generalise this
very much by allowing more general kinds of quantities for your dynamical
variables. They can be any algebraic quantities such that you do not in
general have commutative multiplication (...). Some day people will find the
correct Hamiltonian and there will be some new degrees of freedom, something
which we cannot understand according to classical ideas, playing a role in
the foundations of quantum mechanics." \cite{Dirac1984a}

We insist on the Heisenberg equation and feel free to add certain degrees of freedom that have no counterpart in classical electrodynamics.
Our dynamical variables satisfy the same Heisenberg algebra of canonical commutation relations (CCR) as in the usual quantum field theory, but we work in a representation of CCR that is somewhat unusual. This representation contains {\it four\/} types of additional degrees of freedom, as seen in the arguments of the annihilation operator $a_{\bf b}(\bm R, \bm k, N)$. The degrees of freedom can be given a physical interpretation. ${\bf b}=0,1,2,3$ is a spacetime index, where 1 and 2 correspond to transverse polarizations (massless field of spin 1). 0 and 3 turn out to correspond to two scalar massless fields of dark matter type, unobservable in quantum optical experiments but otherwise probably physical. The wave vector $\bm k$ is not understood as a set of three parameters but as three eigenvalues of some operator. $N$ is a relativistic invariant counting the number of oscillators that form the field --- what is important, $N$ is finite even if the number of different frequencies of the field is infinite, so these are not the oscillators of Heisenberg, Born and Jordan --- one oscillator per mode --- introduced in 1925 in order to quantize electromagnetic fields \cite{HBJ1925}. $R^a=(R^0,\bm R)$ is a timelike world-vector representing, at least tentatively, location of excitations of the oscillators.

The first three sets of quantum numbers were introduced and discussed from various perspectives in a series of papers \cite{I,II,III,CN,V}, and we refer to these representations as the {\it standard-form\/} reducible $N$-representations of CCR. The fact that the representations of CCR are not irreducible agrees with Dirac's intuition, expressed in his last paper \cite{Dirac1984b}, that reducible representations of physical symmetries may be important for field quantization, although our realization of reducibility is not exactly the one Dirac had in mind.

Let us now briefly explain why we claim an extra variable $R^a$ is needed to make the picture complete. Let us consider a null vector $k_a$ that plays a role of a flag-pole for a spinor field $\pi_A(k)$, $k_a=\pi_A(k)\bar\pi_{A'}(k)$ \cite{PR}. $\pi_A(k)$ is given by its flag-pole up to a phase factor, which is essentially where the $U(1)$ gauge group resides (cf. \cite{H}).  Since the transformed spinor field
$
\Lambda\pi_{A}(k) =
\Lambda_{A}^{~~B}\pi_{B}(\Lambda^{-1}k)
$
also satisfies $k_a=\Lambda\pi_A(k)\overline{\Lambda\pi}{_{A'}}(k)$, it follows that $\Lambda\pi_{A}(k)$ and $\pi_{A}(k)$ differ at most by a phase factor: $\Lambda\pi_{A}(k)=e^{-i\Theta(\Lambda,k)}\pi_{A}(k)$. This factor is nothing else but the Wigner phase occurring in massless spin-1/2 unitary representations of the Poincar\'e group \cite{MC99,CW}. Now, the four-potential operator $A_a(x)$ transforms as a four-vector field, simultaneously guaranteeing correct relativistic unitary transformations of momentum-space annihilation operators, provided there exists another spinor field
$\omega_{A}(k)$, $\omega_{A}(k)\pi^{A}(k)=1$, transforming by
\be
\Lambda_{A}^{~~B}\omega_{B}(\Lambda^{-1}k)=e^{+i\Theta(\Lambda,k)}\omega_{A}(k)\label{eq 1}.
\ee
The two null complex vectors, $m_a(k)=\omega_{A}(k)\bar\pi_{A'}(k)$ and its complex conjugate, define circular polarization basis.
Equation of the form (\ref{eq 1}) is fundamental for our formalism, but the problem is that we were not able to find its solution. However, what we managed to do was to find a spin-frame field satisfying
\be
\Lambda_{A}^{~~B}\omega_{B}(\Lambda^{-1}R,\Lambda^{-1}k) &=& e^{+i\Theta(\Lambda,k)}\omega_{A}(R,k),\label{eq 1'}\\
\Lambda_{A}^{~~B}\pi_{B}(\Lambda^{-1}R,\Lambda^{-1}k) &=& e^{-i\Theta(\Lambda,k)}\pi_{A}(R,k),\\
k_a &=& \pi_A(R,k)\bar\pi_{A'}(R,k), \quad \omega_{A}(R,k)\pi^{A}(R,k)=1,
\ee
and that led us to the generalization $a_{\bf b}(\bm R, \bm k, N)$ that we termed the ``center-of-mass" (COM) reducible $N$-representation of CCR. We shall later see that (\ref{eq 1'}) is in many respects similar to the twistor equation \cite{PR2}. Defining circular polarization vectors by $m_a(R,k)=\omega_{A}(R,k)\bar\pi_{A'}(R,k)$ we obtain a simple and elegant formulation of quantum electromagnetic fields.

It remains to test the theory on some meaningful examples. It seems that at this stage an optimal strategy is to discuss exactly solvable models that are known to be infrared-divergent in the usual approaches to field quantization. Following this philosophy, the standard-form $N$-representations were used in \cite{CN} to describe in Heisenberg picture the problem of quantum fields produced by classical pointlike sources. The results were promising but there were some problems with manifest covariance of such a theory, mainly due to the fact that Heisenberg's picture was formulated as an instant-form dynamics \cite{Dirac1949}. In the present paper we replace the instant-form Heisenberg picture by a point-form interaction picture. We treat the fields as being defined not on the entire Minkowski space, but on the Milne universe \cite{Rindler}. Fields are free on the boundary $x_ax^a=0$ of the Milne universe.

We begin, in Section II, with the fundamental dynamical equation for the evolution operator in interaction picture. The corresponding Hamiltonian is constructed from free fields integrated with respect to a Lorentz invariant measure over a hyperboloid of constant proper time $\tau =\sqrt{x_ax^a}$. In Section III we formally solve the Heisenberg equation of motion. We show that although the Hamiltonian is defined for $\tau>0$, the solution is well defined even for $\tau_0=0$. The evolution operator is shown to be given by a coherent-state displacement operator. In Section IV we restrict the analysis to irreducible representations. Taking as an example the case of a pointlike static charge we obtain the Coulomb potential for $\tau>0$, but at $\tau=0$ the potential vanishes. Of course, charge is conserved so vanishing of the potential at the boundary of the Milne universe is a consequence of the initial condition for the Heisenberg dynamics. In Sections V--VII we introduce the standard form of reducible $N$-representation of the four-potential, the corresponding Hilbert space, and the important notion of a vacuum {\it subspace\/}, the set of states annihilated by all annihilation operators. In Section VIII we explicitly construct a representation of the Poincar\'e group in Minkowski space. Although four-translations are not a symmetry group of the Milne universe, they play an important role in definition of free fields at an arbitrary space-time point. This section culminates in the formula for Lorentz transformations of momentum-space operators. It has a characteristic upper-triangular form that mixes all the four spacetime indices of $a_{\bf b}(\bm k, N)$. In particular, the transverse degrees of freedom get mixed with the timelike and longitudinal ones. Section IX contains the central result of the present paper: Mixing disappears in COM representations, i.e. if one replaces $a_{\bf b}(\bm k, N)$ by $a_{\bf b}(\bm R,\bm k, N)$. Now the four spacetime degrees of freedom separate into the direct sum of three massless representations of the Poincar\'e group: Spin 1 (indices 1 and 2), spin 0 (index 3), and spin 0 (index 0). This result would be true even for COM irreducible representations, but we restrict the analysis to the reducible case. Section X shows that the theory reconstructs classical electrodynamics if one defines classical fields by coherent-state averages of Heisenberg-picture operators. We again concentrate on a single pointlike charge moving with constant velocity, and in Section XI we explicitly analyze the resulting potential if the charge is at rest and the vacuum state is spherically symmetric. The reducible representation predicts a potential whose form depends on the choice of a vacuum state. Vacuum is in this representation indeed polarized and screens the pointlike charge by a cloud of effective charge density. The resulting potential depends on the behavior of the vacuum wave-function at infinity and origin (in momentum space). We insist that the wave-function should vanish not only at infinity but also at the origin, since the latter Lorentz invariant boundary condition is needed to regularize infrared divergences in photon statistics. In this case we predict that the effective potential of a pointlike charge decays to zero faster than the Coulomb solution. The photon statistics, discussed in Section XII, is the first prediction where the quantum number $N$ becomes visible: For finite $N$ the resulting probabilities are not Poissonian but of a R\'enyi type, with $\alpha=1-1/N$. The Shannon limit $\alpha\to 1$ simultaneously reconstructs the Poisson statistics and turns the theory into a {\it regularized\/} form of the one known from irreducible representations. In Section XIV we again consider the special case of a pointlike charge. We reconstruct the usual average number of photons typical of Bremsstrahlung but here in an automatically regularized form. We can also split the number of photons into the part typical of acceleration of the charge, and the one corresponding to the inertial part of the trajectory. Both quantities are mathematically well defined.  In Section XIV we discuss in detail which quantities are in our formulation relativistically invariant, and which are covariant. Section XV collects the most essential physical and mathematical differences between what we do and standard quantum field theory. We close the paper by several technical Appendices.

\section{Instant-form and point-form dynamics}

Let $X(0,\bm x)$ be some field at $x_0=0$. Denote by $X(x_0,\bm x)=X(x)=U_0(x_0)^{\dag}X(0,\bm x)U_0(x_0)$ the (instant-form) Heisenberg dynamics of free fields whose Hamiltonian is $H_0$. If the full Hamiltonian is $H=H_0+H_1$ and $U(x_0,y_0)=\exp\big(-i H(x_0-y_0)\big)$ is the full evolution operator, the Heisenberg instant-form dynamics can be written in two equivalent ways
\be
X^H(x)=U(x_0,0)^{\dag}X(0,\bm x)U(x_0,0)=U(x_0,y_0)^{\dag}X^H(y_0,\bm x)U(x_0,y_0)
\ee
and
\be
X^H(x)=U(x_0,0)^{\dag}X(0,\bm x)U(x_0,0)=U_1(x_0)^{\dag}X(x)U_1(x_0)\label{2nd}
\ee
where $U_1(x_0)$ is the solution of the interaction-picture Schr\"odinger equation
\be
i\partial_0 U_1(x_0) &=& H_1(x_0)U_1(x_0),\quad U_1(0)=I,\\
H_1(x_0) &=& U_0(x_0) H_1 U_0(x_0)^{\dag}.
\ee
The interaction Hamiltonian $H_1(x_0)$ is constructed by integrating interaction Hamiltonian density, constructed from {\it free-field\/} operators,  over the hyperplane $\Sigma_x=\{x; x_0={\rm const}\}$. The point $x$ at both sides of (\ref{2nd}) belongs to $\Sigma_x$. If $\Sigma_x$ is an arbitrary space-like hyperplane containing $x$ and belonging to some foliation of a space-time region of interest, one generalizes (\ref{2nd}) as follows (Tomonaga--Schwinger formulation)
\be
X^H(x) &=& U_1(\Sigma_x)^{\dag}X(x)U_1(\Sigma_x)\label{2nd'}\\
i\partial_{\Sigma_x} U_1(\Sigma_x) &=& H_1(\Sigma_x)U_1(\Sigma_x),\quad U_1(\Sigma_0)=I,
\ee
where $H_1(\Sigma_x)$ is obtained by integrating over $\Sigma_x$ the Hamiltonian density constructed from free-field operators evaluated on $\Sigma_x$, and $\partial_{\Sigma_x}$ denotes a derivative with respect to a parameter that labels the hyperplane $\Sigma_x$ in a given foliation (for example $x_0$).

In the Milnean context we employ $\Sigma_x=\{x;\, x^2=\tau^2\geq 0,\, x_0\geq 0\}=:\Sigma_\tau$, $\partial_{\Sigma_x}=d/d\tau$. The Milne universe is the subset ${\cal M}=\bigcup_{\tau\geq 0}\Sigma_\tau$ of the Minkowski space; $\Sigma_\tau$ are regarded as sets of simultaneous events (i.e. the 3D ``space" at ``time" $\tau$).
The Hamiltonian describing interaction of a classical current with quantized electromagnetic fields reads
\be
H_1(\Sigma_\tau) &=& \int d\tilde x_\tau J_a(x_\tau)A^a(x_\tau)=: H_1(\tau)\label{H1 tau}
\ee
where $d\tilde x_\tau=d^3x/\sqrt{1+\bm x^2/\tau^2}$ is the Lorentz invariant measure on $\Sigma_\tau$,  and $x_\tau=(\sqrt{\tau^2+\bm x^2},\bm x)$. The measure is ill defined at $\tau=0$, but the formula $d^4x=d\tau d\tilde x_\tau$ will nevertheless allow us to consider fields with free-field initial conditions on the light cone $\tau=0$, the boundary of the Milne universe. Therefore, we will be able to write the point-form Heisenberg picture dynamics as
\be
X^H(x) &=& U_1(\sqrt{x^2})^{\dag}X(x)U_1(\sqrt{x^2}),\quad U_1(\sqrt{x^2}):=U_1(\sqrt{x^2},0),\quad x\in{\cal M},\label{2nd''}\\
i\frac{d}{d\tau} U_1(\tau,\tau_0) &=& H_1(\tau)U_1(\tau,\tau_0),\quad U_1(\tau_0,\tau_0)=I.\label{H_1''}
\ee
In other words, although the Hamiltonian $H_1(\tau)$ is not well defined at $\tau=0$, the operator $U_1(\tau,\tau_0)$ satisfying (\ref{H_1''}) will be well defined even for $\tau_0=0$. We treat (\ref{2nd''})--(\ref{H_1''}) as the {\it definition\/} of the dynamical system.

\section{Formal algebraic solution}

Consider the Lie $C^*$-algebra of CCR
\be
[a_{\bf a}(\bm \kappa),a_{\bf b}(\bm \kappa')^{\dag}]
&=&
\delta_{\bf ab}\delta(\bm \kappa,\bm \kappa')I(\bm \kappa),\\
{[a_{\bf a}(\bm \kappa),I(\bm \kappa')]} &=& [I(\bm \kappa),a_{\bf a}(\bm \kappa')^{\dag}]=0,\\
\ee
$I(\bm \kappa)^{\dag}=I(\bm \kappa)$,
where ${\bf a},\,{\bf b}=0,\,1,\,2,\,3$, are Minkowski-space indices that can be raised and lowered by means of $g^{\bf ab}=g_{\bf ab}={\rm diag}(1,-1,-1,-1)$. $\delta_{\bf ab}$ is the Kronecker delta and $\delta(\bm \kappa,\bm \kappa')$ is an appropriate Dirac delta distribution. We will consider two types of $\bm \kappa$: $\bm \kappa=\bm k$, where $\bm k$ is a wave vector and
\be
\delta(\bm \kappa,\bm \kappa')
&=&
\delta_0(\bm k,\bm k')=(2\pi)^32|\bm k|\delta^{(3)}(\bm k-\bm k')
\ee
is the invariant measure on the light-cone, and $\bm \kappa=(\bm R,\bm k)$ where $(R_0,\bm R)$ is a future-pointing timelike vector, $R^aR_a=1$. In the latter case
\be
\delta(\bm \kappa,\bm \kappa')
&=&
\delta_0(\bm k,\bm k')\delta_1(\bm R,\bm R')=(2\pi)^62|\bm k|\sqrt{1+\bm R^2}\delta^{(3)}(\bm k-\bm k')\delta^{(3)}(\bm R-\bm R').
\ee
We will discuss two definitions of the electromagnetic four-potential operator. Both involve four polarization degrees of freedom (two transverse, one longitudinal, and one timelike), and a set of continuous degrees of freedom. The first definition, later used in irreducible and standard-form reducible $N$-representations of CCR,  reads
\be
A_{a}(x)
&=& i\int d\tilde k{\,} \left(
  x_{a}(\bm k) a^1(\bm k)
+ y_{a}(\bm k) a^2(\bm k)
+ z_{a}(\bm k) a^3(\bm k)
+ t_{a}(\bm k) a^0(\bm k)^{\dagger} \right) e^{-ik\cdot x}+{\rm H.c}.\\
&=& i\int d\tilde k{\,} \left(
  g_{a}{^1}(\bm k) a_1(\bm k)
+ g_{a}{^2}(\bm k) a_2(\bm k)
+ g_{a}{^3}(\bm k) a_3(\bm k)
+ g_{a}{^0}(\bm k) a_0(\bm k)^{\dagger} \right) e^{-ik\cdot x}+{\rm H.c}.\label{A-first}
\ee
Here $d\tilde k{\,}=d^3k (2\pi)^{-3}(2|\bm k|)^{-1}$ and $x_{a}(\bm k)=g_{a}{_1}(\bm k)$, $\dots$, $t_{a}(\bm k)=g_{a}{_0}(\bm k)$ is a field of Minkowski tetrads \cite{PR}. Details of the notation and properties of the tetrads are explained in the Appendix.
The timelike four-vector $t_{a}(\bm k)$ is accompanied by the {\it creation\/} operator. This allows us to fulfil the covariant commutator
\be
{[A_a(x),A_b(y)]}
&=&
ig_{ab} \hat D(x-y)\label{[A,A]},\\
\hat D(x)
&=&
i\int d\tilde k{\,}I(\bm k)
\big(
e^{-ik\cdot x}
-
e^{ik\cdot x}
\big),
\label{J-P0}
\ee
with $g_{ab}$ being the Minkowski-space metric tensor of signature $(+,-,-,-)$, but --- as we shall see later --- we will not have to use an indefinite-metric formalism, and yet $A_a(x)=A_a(x)^{\dag}$. The hat in $\hat D$ reminds us that the right-hand-side of (\ref{[A,A]}) is not the usual Jordan-Pauli function but a central element of the CCR algebra.

The second definition, important for ``center-of-mass" reducible $N$-representations of CCR, employs certain ambiguity of the number of continuous degrees of freedom. Indeed, let $d\tilde R$ denote a Lorentz invariant measure over a set of parameters $\bm R$. Then
\be
A_{a}(x)
&=& i\int d\tilde k{\,} d\tilde R\left(
  g_{a}{^1}(\bm R,\bm k) a_1(\bm R,\bm k)
+ g_{a}{^2}(\bm R,\bm k) a_2(\bm R,\bm k)
+ g_{a}{^3}(\bm R,\bm k) a_3(\bm R,\bm k)
+ g_{a}{^0}(\bm R,\bm k) a_0(\bm R,\bm k)^{\dagger} \right) e^{-ik\cdot x}+{\rm H.c}\nonumber\\
\label{A-second}
\ee
is a field operator with interesting gauge and covariance properties as we shall see in Section~\ref{COM}. However, calculations involving (\ref{A-second}) are in many respects completely analogous to those with (\ref{A-first}), so for the moment we restrict the analysis to (\ref{A-first}).

Denoting $\hat D_{\rm adv}(x)=-\theta (-x_0)\hat D(x)$ ($\theta$ is the step function) we find the solution of (\ref{H_1''})
\be
U_1(\tau,\tau_0)
&=&
\exp\Big(-i{\int}_{\Sigma_{\tau_0}}^{\Sigma_\tau} d^4xJ^a(x)
A_a(x)\Big)
\exp\Big(-{\frac{i}{2}}\int\int_{\Sigma_{\tau_0}}^{\Sigma_\tau} d^4x_1d^4x_2
J^a(x_1)J_a(x_2)
\hat D_{\rm adv}(x_1-x_2)
\Big)\label{U_1}
\ee
which is well defined also for $\tau_0=0$. The free fields are related to the Heisenberg-picture operators $A_a(x,J)$, $x^2=\tau^2$, by
\be
A_a(x,J)
&=&
U_1(\sqrt{x^2})^{\dag}
A_a(x)
U_1(\sqrt{x^2})
\nonumber\\
&=&
A_a(x)
+
{\int}_{\Sigma_{0}}^{\Sigma_{\tau}} d^4x'
\hat D(x-x')J_a(x').\label{A1}
\ee
At $\tau=0$ we find $U_1(0)=I$ and $A_a^H(x)=A_a(x)$.

Formula (\ref{A1}) shows that $U_1$ is essentially a coherent-state displacement operator, whose general definition in this context should be taken in the form
\be
{\cal D}(\alpha)
&=&
\exp
\int d\tilde k{\,}\Big(
\overline{\alpha_{1}(\bm k)} a^1(\bm k)
+
\overline{\alpha_{2}(\bm k)} a^2(\bm k)
+
\overline{\alpha_{3}(\bm k)} a^3(\bm k)
+
\overline{\alpha_{0}(\bm k)} a^0(\bm k)^{\dagger}
-
{\rm H.c.}
\Big).
\ee
Let us note that again the timelike component $\overline{\alpha_{0}(\bm k)}$ is accompanied by the creation operator. This ensures the correct Minkowskian signature of appropriate bilinear forms, for example,
\be
{\cal D}(\alpha){\cal D}(\beta)
&=&
{\cal D}(\alpha+\beta)
\exp\frac{1}{2}\int d\tilde k{\,}I(\bm k)\Big(
\overline{\alpha_{\bf a}(\bm k)}\beta^{\bf a}(\bm k)
-
\overline{\beta_{\bf a}(\bm k)}\alpha^{\bf a}(\bm k)
\Big).
\ee
The displacement operator shifts the creation and annihilation operators, but leaves the central elements unchanged:
\be
{\cal D}(\alpha)^{\dag}a^{j}(\bm k){\cal D}(\alpha)
&=&
a^{j}(\bm k)
+
\alpha^{j}(\bm k)I(\bm k), \quad j=1,2,3,\\
{\cal D}(\alpha)^{\dag}a^0(\bm k)^{\dag}{\cal D}(\alpha)
&=&
a^0(\bm k)^{\dag}
+
\alpha^{0}(\bm k)I(\bm k),\\
{\cal D}(\alpha)^{\dag}I(\bm k){\cal D}(\alpha)
&=&
I(\bm k).
\ee
In effect the field operators get shifted by central elements,
\be
{\cal D}(\alpha)^{\dag}A_{a}(x){\cal D}(\alpha)
&=&
A_{a}(x)
+
i\int d\tilde k{\,} I(\bm k)
  g_{a}{^{\bf a}}(\bm k) \alpha_{\bf a}(\bm k) e^{-ik\cdot x}+{\rm H.c}.\nonumber\\
&=&
A_{a}(x)
+
\hat A_{a}(\alpha,x).
\ee
In general the shift $\hat A_{a}(\alpha,x)$ is not a classical field but an element of the center of the CCR algebra.

\section{Fields produced by classical currents: Irreducible representations}\label{Fields-irr}

In irreducible representations all central elements are proportional to an identity operator $I$, i.e. $I(\bm k)=ZI$, with some constant $Z>0$. We assume there exists a vacuum state $|0\rangle$ annihilated by all $a_{\bf a}(\bm k)$. (The apparently more general form $I(\bm k)=Z(\bm k)I$, where $Z(\bm k)$ is a function, would lead to inconsistencies with the Lorentz covariance of the theory; a kind of $Z$ and $Z(\bm k)=\langle 0|I(\bm k)|0\rangle$ will however appear in the reducible $N$-representations, and this is why we do not put here $Z=1$ but keep it general.) The solutions are
\be
U_1(\tau,\tau_0)
&=&
\exp\Big(-i{\int}_{\Sigma_{\tau_0}}^{\Sigma_\tau} d^4xJ^a(x)
A_a(x)\Big)
\exp\Big(-{\frac{i}{2}}\int\int_{\Sigma_{\tau_0}}^{\Sigma_\tau} d^4x_1d^4x_2
ZJ^a(x_1)J_a(x_2)
D_{\rm adv}(x_1-x_2)
\Big)\label{U_1-i}\\
A_a(x,J)
&=&
A_a(x)
+
ZI{\int}_{\Sigma_{0}}^{\Sigma_{\tau}} d^4x'
D(x-x')J{_a}(x').\label{A1-i}
\ee
Here $D$ and $D_{\rm adv}$ are the ordinary classical expressions. Eq.~(\ref{U_1-i}) means that for $Z\neq 1$ the effective {\it physically observable\/} current (the one responsible for photon statistics) is not $J^a(x)$ but $J_{\rm ren}^a(x)=Z^{1/2}J_a(x)$, and thus the appropriate physical fields are not $A_a(x)$ and $A_a(x,J)$, but rather $A_{\rm ren}^a(x)=Z^{-1/2}A^a(x)$ and $A_a{_{\rm ren}}(x,J)=Z^{-1/2}A_a(x,J)$.  Now consider the coherent state $|\alpha\rangle={\cal D}(\alpha)|0\rangle$. A classical field corresponding to $A_a(x,J)$ is the average
\be
\langle\alpha|A_a{_{\rm ren}}(x,J)|\alpha\rangle
&=&
i\int d\tilde k{\,}  g_{a}{_{\bf a}}(\bm k) \alpha^{\bf a}_{\rm ren}(\bm k) e^{-ik\cdot x}+{\rm c.c}.
\nonumber\\
&\pp=&
+
{\int}_{\Sigma_{0}}^{\Sigma_{\tau}} d^4x'
D(x-x')J{_{\rm ren}}{_a}(x'),\label{A-aver}
\ee
where $\alpha^{\bf a}_{\rm ren}(\bm k)=Z^{1/2}\alpha^{\bf a}(\bm k)$. The latter is easier to understand if one notices that
$a^{\bf a}(\bm k)|\alpha\rangle=Z\alpha^{\bf a}(\bm k)|\alpha\rangle$. In other words, for $Z\neq 1$ the physical fields are
$a^{\bf a}_{\rm ren}(\bm k)=Z^{-1/2}a^{\bf a}(\bm k)$ satisfying $a^{\bf a}_{\rm ren}(\bm k)|\alpha\rangle=\alpha^{\bf a}_{\rm ren}(\bm k)|\alpha\rangle$ and
\be
[a^{\bf a}_{\rm ren}(\bm k),a^{\bf b}_{\rm ren}(\bm k')^{\dag}]
&=&
\delta^{\bf ab}\delta_0(\bm k,\bm k')I.
\ee
So effectively we are back to the representations with $Z=1$. We could, of course, renormalize the representation from the very beginning by dividing both sides of CCR by $Z$. We decided not to do so, however, since when it comes to the reducible $N$-representations the procedure of identifying physical quantities is not that obvious. We will see that at the end the correct formalism will effectively require renormalization equivalent to dividing both sides of CCR by some finite, Lorentz invariant $Z$, but this $Z$ will have a different meaning and origin than the constant occurring in irreducible representations.

We know that for pointlike sources the operator $U_1(\tau)$ does not exist (in irreducible representations!) due to infrared divergences. It is nevertheless interesting to check what are the implications for (\ref{A-aver}) of the fact that one integrates from $\tau=0$ and not from $\tau=-\infty$. Let the curve $\tau\mapsto X_a(\tau)=\big(\sqrt{\tau^2+\bm X(\tau)^2},\bm X(\tau)\big)$ describe the trajectory of a pointlike charge $q$.
The current is
\be
J^a(x) &=& q \int_{-\infty}^{\infty}ds \frac{dX^a(s)}{ds}\delta^{(4)}(x-X(s))
\label{J point}
\ee
The current is conserved also in the Milnean framework as a trivial consequence of its conservation in Minkowski space.
Now that we know how to deal with $Z\neq 1$ we can renormalize and set $Z=1$, so that $q=q_{\rm ren}$. Evaluating the space-like integral we obtain
\be
A_a(x,J)
&=&
A_a(x)
+
q
\int_{0}^{\sqrt{x^2}}ds \frac{dX_a(s)}{ds}
D(x-X(s)).
\ee
Let us now concentrate on the Coulomb field, i.e. $X_{\bf a}(s)=(s,0,0,0)$ and vacuum initial condition. Then only $\langle 0|A_0(x,J)|0\rangle$ is nonvanishing.
Performing standard computations and taking into account the signs of the arguments,
\be
x_0-\tau+|\bm x|
&=&
\sqrt{\tau^2+\bm x^2}-\tau +|\bm x|\geq 0,\nonumber\\
x_0-\tau-|\bm x|
&=&
\sqrt{\tau^2+\bm x^2}-(\tau +|\bm x|)
\leq 0,\nonumber\\
x_0\pm|\bm x| &\geq& 0
\ee
we arrive at
\be
\langle 0|A_0(x,J)|0\rangle
&=&
\frac{q}{(2\pi)^2|\bm x|}
\int_0^\infty dk
\Big(
\frac{\sin k(x_0-\tau+|\bm x|)+\sin k(|\bm x|-x_0+\tau)}{k}
-
\frac{\sin k(x_0+|\bm x|)-\sin k(x_0-|\bm x|)}{k}
\Big)
\nonumber\\
&=&
\left\{
\begin{array}{cl}
0 &{\rm for}\quad \tau=0\\
\frac{q}{4\pi|\bm x|} &{\rm for}\quad \tau>0
\end{array}
\right.
.
\ee
We conclude that observations performed at the cosmic time $\tau>0$ are unable to tell the difference between Coulomb fields corresponding to the Minkowski space and those typical of the Milne universe. The situation will change if one quantizes the field in an $N$-representation.

\section{Standard form of reducible $N<\infty$ representations of CCR}\label{Sec N<}

What we call in this paper a standard form of $N$-representations of CCR was introduced in the context of nonrelativistic quantum optics in \cite{I}. Further elements of the construction (coherent states, Poincar\'e group) were discussed in \cite{II}, a fermionic (canonical anti-commutation relations) were introduced in \cite{III}, and the representation which is essentially the one we discuss in the present section was given in \cite{CN}. A preliminary comparison of $N$-representations with experimental cavity QED can be found in \cite{V}.

One begins with four operators, $a_0$, $a_1$, $a_2$, $a_3$,
satisfying commutation relations typical of an
{\it irreducible\/} representation of CCR:
$[a_{\bf a},a_{\bf b}^{\dag}]=\delta_{\bf ab}1$.
Let $|0\rangle$ denote their common vacuum,
i.e.~$a_{\bf a}|0\rangle=0$.
Now take the kets $|\bm k\rangle$ normalized with respect
to the light-cone delta function
\be
\langle\bm k|\bm k'\rangle
&=&
\delta_0(\bm k,\bm k')
=
(2\pi)^3 2|\bm k|\delta^{(3)}(\bm k-\bm k').
\ee
What we call the $N=1$ (or 1-oscillator) representation
of CCR acts in the Hilbert space
${\cal H}(1)$ spanned by kets of the form
\be
|\bm k,n_0,n_1,n_2,n_3\rangle
=
|\bm k\rangle\otimes
\frac{(a_0^{\dag})^{n_0}(a_1^{\dag})^{n_1}(a_2^{\dag})^{n_2}(a_3^{\dag})^{n_3}}
{\sqrt{n_0!n_1!n_2!n_3!}}|0\rangle.\nonumber
\ee
Physically, ${\cal H}(1)$ may be regarded as representing the space
of states of a single four-dimensional oscillator.
The 1-oscillator representation is defined by
\be
a_{\bf a}(\bm k,1)=|\bm k\rangle\langle\bm k|\otimes a_{\bf a}.
\ee
This representation is reducible since the commutator
\be
[a_{\bf a}(\bm k,1),a_{\bf b}(\bm k',1)^{\dag}]
&=&
\delta_{\bf ab}\delta_0(\bm k,\bm k')
|\bm k\rangle\langle\bm k|\otimes 1
\ee
involves at the right-hand-side the operator-valued distribution
$I(\bm k,1)=|\bm k\rangle\langle\bm k|\otimes 1$ belonging to
the center of the algebra,
$[a_{\bf a}(\bm k,1),I(\bm k',1)]=[I(\bm k,1),a_{\bf a}(\bm k',1)^{\dag}]=0$,
for all $\bm k$, $\bm k'$, ${\bf a}$.
Operators
$I(\bm k,1)$ form a resolution of unity
\be
\int d\tilde k{\,}I(\bm k,1)
&=&
\int d\tilde k{\,}|\bm k\rangle\langle\bm k|\otimes 1=I\otimes 1=I(1).
\ee
Here $I(1)$ is the identity operator in ${\cal H}(1)$.

For arbitrary $N$ the representation is constructed as follows.
Define
\be
{\cal H}(N)=\underbrace{{\cal H}(1)\otimes\dots\otimes {\cal H}(1)}_N
\ee
and let $A$ be an arbitrary operator defined for $N=1$. Let
\be
A^{(n)}=\underbrace{I(1)\otimes\dots\otimes I(1)}_{n-1}\otimes A
\otimes \underbrace{I(1)\otimes\dots\otimes I(1)}_{N-n}.
\ee
The $N$ oscillator extension of $a_{\bf a}(\bm k,1)$ is defined by
\be
a_{\bf a}(\bm k,N)
&=&
\frac{1}{\sqrt{N}}\sum_{n=1}^N a_{\bf a}(\bm k,1)^{(n)}
\ee
and satisfies the reducible representation
\be
[a_{\bf a}(\bm k,N),a_{\bf b}(\bm k',N)^{\dag}]
&=&
\delta_{\bf ab}\delta_0(\bm k,\bm k')
I(\bm k,N)
\ee
where
\be
I(\bm k,N)
&=&
\frac{1}{N}\sum_{n=1}^N I(\bm k,1)^{(n)}.\label{uu I(k)}
\ee
As before we find the resolution of unity
\be
\int d\tilde k{\,}I(\bm k,N)
&=&
I(N)
\ee
where $I(N)$ is the identity operator in ${\cal H}(N)$.

\section{Vacuum, multiphoton, and coherent states in standard-form $N$-representations}

Vacuum is in this representation any state annihilated by all annihilation operators. Such a state is neither unique nor Poincar\'e invariant --- we will later show, when we introduce ``center-of-mass" $N$-representations, how to construct a representation of the Poincar\'e group in a way that ensures invariance of the entire {\it subspace\/} of vacuum states. The set of vacuum states is spanned by vectors of the form
\be
|\bm k_1,0,0,0,0\rangle\otimes\dots\otimes |\bm k_N,0,0,0,0\rangle.
\ee
Now let $|0,1\rangle=\int d\tilde k{\,}O(\bm k)|\bm k,0,0,0,0\rangle$, $\int d\tilde k{\,}|O(\bm k)|^2=1$, be a vacuum state for $N=1$. For arbitrary $N$ a vacuum is defined as
\be
|0,N\rangle=\underbrace{|0,1\rangle\otimes\dots\otimes|0,1\rangle}_N.
\ee
Acting on a vacuum with the $N$-representation of the displacement operator
\be
{\cal D}(\alpha,N)
&=&
\exp
\int d\tilde k{\,}\Big(
\overline{\alpha_{1}(\bm k)} a^1(\bm k,N)
+
\overline{\alpha_{2}(\bm k)} a^2(\bm k,N)
+
\overline{\alpha_{3}(\bm k)} a^3(\bm k,N)
+
\overline{\alpha_{0}(\bm k)} a^0(\bm k,N)^{\dagger}
-
{\rm H.c.}
\Big)
\ee
we obtain a coherent state
\be
|\alpha,N\rangle={\cal D}(\alpha,N)|0,N\rangle.
\ee
A a linear combination of vectors of the form
\be
|\bm k_1,\dots,\bm k_N, n_0^{(1)},n_1^{(1)},n_2^{(1)},n_3^{(1)},\dots,n_1^{(N)},n_2^{(N)},n_3^{(N)}\rangle
=
|\bm k_1,n_0^{(1)},n_1^{(1)},n_2^{(1)},n_3^{(1)}\rangle\otimes\dots\otimes |\bm k_N,n_0^{(N)},n_1^{(N)},n_2^{(N)},n_3^{(N)}\rangle.\label{n-fot}
\ee
is regarded as a state of $n_3=\sum_{j=1}^Nn_3^{(j)}$ longitudinal and $n_0=\sum_{j=1}^Nn_0^{(j)}$ timelike ``photons"; $n_1=\sum_{j=1}^Nn_1^{(j)}$ and $n_2=\sum_{j=1}^Nn_2^{(j)}$ describe the numbers of transverse photons. Timelike and longitudinal ``photons" are, by assumption, unobservable in quantum optical measurements. Number of photons is therefore identified with the number of excitations in the $N$-oscillator system. In particular, an $n$th power of a creation operator acting on a vacuum state creates an $n$-photon state and coherent states have a Poisson statistics. We shall later discuss these issues on the explicit example of the photon statistics of fields produced by classical sources.

\section{Standard-form $N<\infty$ representation of the four-potential operator}

The potential operator in this representation reads
\begin{eqnarray}
A_{a}(x,N)
&=& i\int d\tilde k{\,}\left(
  g_{a}^{~1}(\bm k)a_1(\bm k,N)
+ g_{a}^{~2}(\bm k)a_2(\bm k,N)
+ g_{a}^{~3}(\bm k)a_3(\bm k,N)
+ g_{a}^{~0}(\bm k)a_0(\bm k,N)^{\dagger}
\right) e^{-ik\cdot x}+{\rm H.c.}
\end{eqnarray}
Let us note that defining
$\hat k^a={\textstyle\int} d\tilde k{\,}k^a|\bm k\rangle\langle\bm k|$
we can write the $N=1$ case as
\be
A_{a}(x,1)
&=& i
\left(
  g_{a}^{~1}(\hat{\bm k})e^{-i\hat k\cdot x}\otimes a_1
+ g_{a}^{~2}(\hat{\bm k})e^{-i\hat k\cdot x}\otimes a_2
+ g_{a}^{~3}(\hat{\bm k})e^{-i\hat k\cdot x}\otimes a_3
+ g_{a}^{~0}(\hat{\bm k})e^{-i\hat k\cdot x}\otimes a_0^{\dagger}
\right) +{\rm H.c.}
\ee
where all the integrals have been absorbed into spectral decompositions of $\hat k^a$.
The commutator
\be
{[A_{a}(x,N),A_{b}(y,N)]}
&=&
ig_{ab} \hat D(x-y,N)\label{[A,A]'}
\ee
involves the operator analogue of the Jordan-Pauli function
\be
\hat D(x,N)
&=&
i\int d\tilde k{\,}I(\bm k,N)
\big(
e^{-ik\cdot x}
-
e^{ik\cdot x}
\big).
\label{J-P}
\ee
As before, the correct signature of the metric tensor in (\ref{[A,A]'})
comes from the Bogoliubov-type structure of the positive-frequency
part of $A_a(x,N)$, i.e. the combination of annihilation and
creation operators. If one had replaced $a_0^{\dag}$ by $a_0$
one would have been forced to depart either from positivity of
the scalar product or unitarity of evolution.

\section{Representation of the Poincar\'e group in Minkowski background space}

We first construct a standard-form representation of an operator $U(\Lambda,y,1)$ (i.e. $N=1$) acting by
\be
U(\Lambda,y,1)^{\dag}A_a(x,1)U(\Lambda,y,1)=\Lambda{_a}{^b}
A_b\big(\Lambda^{-1}(x-y),1\big),
\label{PG1}
\ee
and then extend it to arbitrary $N$ by
\be
U(\Lambda,y,N)
&=&\underbrace{U(\Lambda,y,1)\otimes\dots\otimes U(\Lambda,y,1)}_N,\label{PGN}\\
U(\Lambda,y,N)^{\dag}A_a(x,N)U(\Lambda,y,N) &=&
\Lambda{_a}{^b}A_b\big(\Lambda^{-1}(x-y),N\big).
\ee
\subsection{Four-translations}

The 4-momentum for $N=1$ reads
\be
P_a(1)
&=&
{\textstyle\int} d\tilde k{\,}k_a|\bm k\rangle\langle \bm k|\otimes
\big(a^{\dag}_1a_1+a^{\dag}_2a_2+a^{\dag}_3a_3-a^{\dag}_0a_0\big)
\\
&=&
-{\textstyle\int} d\tilde k{\,}k_a|\bm k\rangle\langle \bm k|\otimes
a^{\dag}_{\bf a}a^{\bf a}\label{P}\\
&=&
{\textstyle\int} d\tilde k{\,}k_a
\big(
\hat n_1(\bm k,1)
+
\hat n_2(\bm k,1)
+
\hat n_3(\bm k,1)
-
\hat n_0(\bm k,1)
\big).
\label{Pn}
\ee
The form (\ref{Pn}) defines four number operators $\hat n_{\bf a}(\bm k,1)$.
One immediately verifies that
\be
e^{iP(1)\cdot x}a_1(\bm k,1)e^{-iP(1)\cdot x} &=& a_1(\bm k,1) e^{-ix\cdot k},\\
e^{iP(1)\cdot x}a_2(\bm k,1)e^{-iP(1)\cdot x} &=& a_2(\bm k,1) e^{-ix\cdot k},\\
e^{iP(1)\cdot x}a_3(\bm k,1)e^{-iP(1)\cdot x} &=& a_3(\bm k,1) e^{-ix\cdot k},\\
e^{iP(1)\cdot x}a_0(\bm k,1)^{\dag}e^{-iP(1)\cdot x} &=& a_0(\bm k,1)^{\dag} e^{-ix\cdot k},
\ee
implying
\be
U(\bm 1,y,1)^{\dag}A_{a}(x,1)U(\bm 1,y,1) &=&
A_{a}(x-y,1).
\ee
The 4-momentum for arbitrary $N$ reads
\be
P_a(N) &=&\textstyle{\sum}_{n=1}^N P_a(1)^{(n)}\label{uu P}\\
&=&
{\textstyle\int} d\tilde k{\,}k_a
\big(
\hat n_1(\bm k,N)
+
\hat n_2(\bm k,N)
+
\hat n_3(\bm k,N)
-
\hat n_0(\bm k,N)
\big).
\label{PnN}
\ee
The number operators, defined by (\ref{PnN}), satisfy
\be
\hat n_{\bf a}(\bm k,N)
&=&
\textstyle{\sum}_{n=1}^N
\hat n_{\bf a}(\bm k,1)^{(n)}.
\ee
We again find the correct formula
\be
e^{iP(N)\cdot x}a_1(\bm k,N)e^{-iP(N)\cdot x} &=& a_1(\bm k,N) e^{-ix\cdot k},\\
e^{iP(N)\cdot x}a_2(\bm k,N)e^{-iP(N)\cdot x} &=& a_2(\bm k,N) e^{-ix\cdot k},\\
e^{iP(N)\cdot x}a_3(\bm k,N)e^{-iP(N)\cdot x} &=& a_3(\bm k,N) e^{-ix\cdot k},\\
e^{iP(N)\cdot x}a_0(\bm k,N)^{\dag}e^{-iP(N)\cdot x} &=& a_0(\bm k,N)^{\dag} e^{-ix\cdot k},
\ee
implying
\be
U(\bm 1,y,N)^{\dag}A_{a}(x,N)U(\bm 1,y,N) &=&
A_{a}(x-y,N).
\ee
The form (\ref{uu P}) is characteristic of a 4-momentum of $N$
non-interacting particles. These particles (four-dimensional oscillators)
have no counterpart in classical electrodynamics.

Vectors (\ref{n-fot}) are simultaneously the eigenvectors of $P_a(N)$,
\be
{}&{}&P^a(N)
|\bm k_1,\dots,\bm k_N, n_0^{(1)},\dots,n_3^{(N)}\rangle
\nonumber\\
&{}&\pp==
\Big(k^a_1\big(n_1^{(1)}+n_2^{(1)}+n_3^{(1)}-n_0^{(1)}\big)+\dots+k^a_N\big(n_1^{(N)}+n_2^{(N)}+n_3^{(N)}-n_0^{(N)}\big)\Big)
|\bm k_1,\dots,\bm k_N, n_0^{(1)},\dots,n_3^{(N)}\rangle
\ee

\subsection{Boosts and rotations}

We begin with the generators
\be
J_{1} &=& i(a_{3}^{\dagger}a_{2}-a_{2}^{\dagger}a_{3}),\label{J_1}\\
J_{2} &=& i(a_{1}^{\dagger}a_{3}-a_{3}^{\dagger}a_{1}),\\
J_{3} &=& i(a_{2}^{\dagger}a_{1}-a_{1}^{\dagger}a_{2}),\\
K_{1} &=& i(a_{0}^{\dagger}a_{1}^{\dagger}-a_{0}a_{1}),\\
K_{2} &=& i(a_{0}^{\dagger}a_{2}^{\dagger}-a_{0}a_{2}),\\
K_{3} &=& i(a_{0}^{\dagger}a_{3}^{\dagger}-a_{0}a_{3}),\label{K_3}\\
{[J_{i},J_{j}]} &=& i\varepsilon_{ijk}J_{k},\\
{[K_{i},K_{j}]} &=& -i\varepsilon_{ijk}J_{k},\\
{[J_{i},K_{j}]} &=& i\varepsilon_{ijk}K_{k},
\ee
of SO(1,3). Of particular importance is the combination $b=a_{3}-a_{0}^{\dagger}$, $[b,b^{\dag}]=0$, occurring in the generators of E(2)
\be
L_{1} &=& J_{1}+K_{2}=i( b^{\dagger}a_{2}-a_{2}^{\dagger}b),\\
L_{2}
&=& J_{2}-K_{1}=i(a_{1}^{\dagger}b-b^{\dagger}a_{1}),\\
L_{3} &=& J_{3},\\
{[L_{3},L_{1}]} &=& iL_{2},\\
{[L_{2},L_{3}]} &=& iL_{1},\\
{[L_{1},L_{2}]} &=& 0.
\ee
We prove in the Appendix that for $N=1$ the representation is
\be
U(\Lambda,0,1) &=&\int d\tilde k{\,}|\bm k\rangle\langle
\bm{\Lambda^{-1}k}|\otimes
e^{-i|\phi(\Lambda,\bm k)|\sin \xi(\Lambda,\bm k)L_1}
e^{-i|\phi(\Lambda,\bm k)|\cos \xi(\Lambda,\bm k)L_2}
e^{-i 2\Theta(\Lambda,\bm k) L_3}.\label{sf U}
\ee
Here $\phi(\Lambda,\bm k) = |\phi(\Lambda,\bm k)| e^{i\xi(\Lambda,\bm k)}$ and $\Theta(\Lambda,\bm k)$ are related to the spin-frame field (see Appendix)  $\omega^{A}(\bm k)$, $\pi^{A}(\bm k)$, $\omega_{A}(\bm k)\pi^{A}(\bm k)=1$, $k^a=\pi^{A}(\bm k)\bar\pi^{A'}(\bm k)$, by
\be
e^{i\Theta(\Lambda,\bm k)}
&=&
\pi^{A}(\bm k)
\Lambda\omega_{A}(\bm k),\\
\phi(\Lambda,\bm k)
&=&
e^{-i\Theta(\Lambda,\bm k)}\omega_{A}(\bm k)
\Lambda\omega^{A}(\bm k),\\
\Lambda\pi_{A}(\bm k) &=&
\Lambda_{A}^{~~B}\pi_{B}(\boldsymbol{\Lambda^{-1}k})=e^{-i\Theta(\Lambda,\bm k)}\pi_{A}(\bm k),\\
\Lambda\omega_{A}(\bm k) &=&
\Lambda_{A}^{~~B}\omega_{B}(\boldsymbol{\Lambda^{-1}k}).
\ee
$\Theta(\Lambda,\bm k)$ is the Wigner phase, and $\phi(\Lambda,\bm k)$ is a quantity of a gauge type that has no geometric meaning (due to the ambiguity in the definition of $\omega^{A}(\bm k)$, see Appendix). Later, in ``center-of-mass" representations, we shall find $\phi=0$.

The required transformation rule
\be
U(\Lambda,0,1)^{\dag}A_a(x,1)U(\Lambda,0,1)=\Lambda{_a}{^b}
A_b\big(\Lambda^{-1}x,1\big)
\ee
extends to arbitrary $N$ by (\ref{PGN}). The momentum-space transformations have a triangular form if expressed in terms of the combinations $a_{\pm}(\bm k,N)=(a_{1}(\bm k,N)\pm ia_{2}(\bm k,N))/\sqrt{2}$ (circular polarizations) and $b_\pm(\bm k,N)=(a_{3}(\bm k,N)\pm a_{0}(\bm k,N)^{\dag})/\sqrt{2}$,
\be
\left(
\begin{array}{c}
U(\Lambda,0,N)^{\dag}b_+(\bm k,N)U(\Lambda,0,N)\\
U(\Lambda,0,N)^{\dag}a_+(\bm k,N)U(\Lambda,0,N)\\
U(\Lambda,0,N)^{\dag}a_-(\bm k,N)U(\Lambda,0,N)\\
U(\Lambda,0,N)^{\dag}b_-(\bm k,N)U(\Lambda,0,N)
\end{array}
\right)
&=&
\left(
\begin{array}{cccc}
1 & -\phi(\Lambda,\bm k)e^{2i\Theta(\Lambda,\bm k)} & -\bar\phi(\Lambda,\bm k)e^{-2i\Theta(\Lambda,\bm k)} & -|\phi(\Lambda,\bm k)|^2\\
0 & e^{2i\Theta(\Lambda,\bm k)} & 0 & \bar\phi(\Lambda,\bm k)\\
0 & 0 & e^{-2i\Theta(\Lambda,\bm k)} & \phi(\Lambda,\bm k)\\
0 & 0 & 0 & 1
\end{array}
\right)
\left(
\begin{array}{c}
b_+(\bm{\Lambda^{-1}k},N)\\
a_+(\bm{\Lambda^{-1}k},N)\\
a_-(\bm{\Lambda^{-1}k},N)\\
b_-(\bm{\Lambda^{-1}k},N)
\end{array}
\right)\nonumber\\
\label{APM}
\ee
Comparing (\ref{APM}) with Eqs.~(74)--(76) from \cite{CN} we see that the former reduce to the latter if $\phi=0$, a consequence of the fact that (74)--(76) from \cite{CN} change gauge, which has yet to be corrected by Eq.~(40) in \cite{CN}. Our present formalism takes care of this gauge correction, but in consequence the momentum-space circular-polarization annihilation operators do not transform as irreducible spin-1 zero-mass representations of the Poincar\'e group. In the next section we discuss a remedy to this difficulty.

\section{``Center-of-mass" $N$-representations of CCR}\label{COM}

In non-relativistic quantum mechanics the algebra $[a_k,a_l^{\dag}]=\delta_{kl}$, occurring for a harmonic three-dimensional oscillator, corresponds to the relative coordinate $\bm r=\bm x_1-\bm x_2$. Our harmonic oscillators are characterized by quantum numbers that index the basis: $|\bm k,n_0,n_1,n_2,n_3\rangle$. The four numbers $n_{\bf a}$ describe excitations of some four-dimensional ``relative coordinate", analogous to $\bm r$. The  quantum numbers $\bm k$ replace the {\it parameter\/} $\omega$ characterizing the non-relativistic potential $m\omega^2\bm r^2/2$. Now, we know that physical oscillators are characterized also by the center-of-mass coordinate $\bm R$. Can and should we introduce such a type of additional degree of freedom in our reducible representations of CCR?

It turns out that there is one place in the formalism where the presence of a timelike and independent of $\bm k$ world-vector $R^a$ would be of some help. In order to understand it let us return to the function $\phi(\Lambda,\bm k)$ we have encountered in the previous section. The formalism would be much more elegant if we could automatically guarantee $\phi(\Lambda,\bm k)=0$. This is equivalent to
\be
\Lambda\omega_{A}(\bm k) &=&
\Lambda_{A}^{~~B}\omega_{B}(\boldsymbol{\Lambda^{-1}k})
=
e^{i\Theta(\Lambda,\bm k)}
\omega_{A}(\bm k).
\ee
The problem is that we have not managed to find a spin-frame with this property. However, the spin-frame, with $\bm k$-independent $\nu^A$ and $R^a$,
\be
\pi^A(\bm R,\bm k)
&=&
\frac{k^{AA'}\bar \nu_{A'}}
{\sqrt{k^{BB'}\nu_B
\bar \nu_{B'}}}=\pi^A(\bm k),\label{pppp}\\
\omega^{A}(\bm R,\bm k)
&=&-
\frac{R^{AA'}\bar \pi_{A'}(\bm k)}
{R^ak_a},
\label{oooo}
\ee
satisfies all the requirements we have imposed on the spin-frames so far, plus
\be
\Lambda\omega_{A}(\bm R,\bm k) &=&
\Lambda_{A}^{~~B}\omega_{B}(\bm{\Lambda^{-1}R},\bm{\Lambda^{-1}k})
=
e^{i\Theta(\Lambda,\bm k)}
\omega_{A}(\bm R,\bm k).
\ee
As usual, by $\bm{\Lambda^{-1}R}$ we denote a spacelike component of $\Lambda^{-1}{_a}{^b}R_b$. $R^a$ can be in principle timelike, null, or spacelike, but only for a timelike $R^a$ the denominator in (\ref{oooo}) is never vanishing. The zero-homogeneity $\omega_{A}(\lambda\bm R,\bm k)=
\omega_{A}(\bm R,\bm k)$ implies that one can assume $R^aR_a=1$.
So take the kets $|\bm R\rangle$ normalized with respect
to the delta function,
\be
\langle\bm R|\bm R'\rangle
&=&
\delta_1(\bm R,\bm R')
=
(2\pi)^3 2\sqrt{1+\bm R^2}\delta^{(3)}(\bm R-\bm R').
\ee
The $N=1$ ``center-of-mass" (COM) representation
of CCR acts in the Hilbert space
${\cal H}(1)$ spanned by kets of the form
\be
|\bm R,\bm k,n_0,n_1,n_2,n_3\rangle
=
|\bm R,\bm k\rangle\otimes
\frac{(a_0^{\dag})^{n_0}(a_1^{\dag})^{n_1}(a_2^{\dag})^{n_2}(a_3^{\dag})^{n_3}}
{\sqrt{n_0!n_1!n_2!n_3!}}|0\rangle,\nonumber
\ee
$|\bm R,\bm k\rangle=|\bm R\rangle\otimes|\bm k\rangle$.
The 1-oscillator COM representation is defined by
\be
a_{\bf a}(\bm R,\bm k,1) &=&|\bm R,\bm k\rangle\langle\bm R,\bm k|\otimes a_{\bf a},\\
{[a_{\bf a}(\bm R,\bm k,1),a_{\bf b}(\bm R',\bm k',1)^{\dag}]}
&=&
\delta_{\bf ab}\delta(\bm R,\bm k,\bm R',\bm k')I(\bm R,\bm k,1),\\
\delta(\bm R,\bm k,\bm R',\bm k')
&=&
\delta_1(\bm R,\bm R')\delta_0(\bm k,\bm k'),\\
I(\bm R,\bm k,1) &=&
|\bm R,\bm k\rangle\langle\bm R,\bm k|\otimes 1.
\ee
Operators
$I(\bm R,\bm k,1)$ form a resolution of unity
\be
\int d\tilde k{\,}d\tilde R{\,} I(\bm R,\bm k,1)
&=&
I\otimes I\otimes 1
=I(1),\label{I(1)-R}
\ee
where $d\tilde R$ is the invariant measure on the hyperboloid $R^aR_a=1$. The two identity operators $I$ in (\ref{I(1)-R}) act in different Hilbert spaces but to simplify notation we denote them by the same symbol.
For arbitrary $N$ we proceed as before
\be
{\cal H}(N)=\underbrace{{\cal H}(1)\otimes\dots\otimes {\cal H}(1)}_N
\ee
The $N$ oscillator extension of $a_{\bf a}(\bm R,\bm k,1)$ is defined by
\be
a_{\bf a}(\bm R,\bm k,N)
&=&
\frac{1}{\sqrt{N}}\sum_{n=1}^N a_{\bf a}(\bm R,\bm k,1)^{(n)}
\ee
and satisfies the reducible representation
\be
[a_{\bf a}(\bm R,\bm k,N),a_{\bf b}(\bm R',\bm k',N)^{\dag}]
&=&
\delta_{\bf ab}\delta(\bm R,\bm k,\bm R',\bm k')I(\bm R,\bm k,N),\\
I(\bm R,\bm k,N)
&=&
\frac{1}{N}\sum_{n=1}^N I(\bm R,\bm k,1)^{(n)}.\label{uu I(k)-R}
\ee
As before we find the resolution of unity
$
\int d\tilde k{\,}d\tilde R{\,}I(\bm R,\bm k,N)
=
I(N)
$
where $I(N)$ is the identity operator in ${\cal H}(N)$.
The potential operator in the new representation is defined in exact analogy to our previous definitions,
\be
A_{a}(x,N)
&=& i\int d\tilde k{\,}d\tilde R\left(
  g_{a}^{~1}(\bm R,\bm k)a_1(\bm R,\bm k,N)
+ \dots + g_{a}^{~0}(\bm R,\bm k)a_0(\bm R,\bm k,N)^{\dagger}
\right) e^{-ik\cdot x}+{\rm H.c.}
\ee
The Minkowski tetrad $g_{a}{^{\bf a}}(\bm R,\bm k)$ is linked to the spin frame by (\ref{4}).
The analogue of the Jordan-Pauli function is
\be
\hat D(x,1)
&=&
i\int d\tilde k{\,}d\tilde R{\,}I(\bm R,\bm k,1)
\big(
e^{-ik\cdot x}
-
e^{ik\cdot x}
\big)
=
I\otimes i\int d\tilde k{\,}I(\bm k,1)
\big(
e^{-ik\cdot x}
-
e^{ik\cdot x}
\big),
\label{J-P-R}\\
\hat D(x,N)
&=&
i\int d\tilde k{\,}d\tilde R{\,}I(\bm R,\bm k,N)
\big(
e^{-ik\cdot x}
-
e^{ik\cdot x}
\big)
=
\frac{1}{N}\sum_{n=1}^N \hat D(x,1)^{(n)}.
\ee
The 4-momentum for $N=1$ reads
\be
P_a(1)
&=&
-I\otimes {\textstyle\int} d\tilde k{\,}k_a|\bm k\rangle\langle \bm k|\otimes
a^{\dag}_{\bf a}a^{\bf a}
\label{P-R}\\
&=&
{\textstyle\int} d\tilde k{\,}k_a
\big(
\hat n_1(\bm k,1)
+
\hat n_2(\bm k,1)
+
\hat n_3(\bm k,1)
-
\hat n_0(\bm k,1)
\big),
\label{Pn-R}
\ee
and for arbitrary $N$ we employ the extension (\ref{uu P}),
\be
P_a(N)
&=&
{\textstyle\int} d\tilde k{\,}k_a
\big(
\hat n_1(\bm k,N)
+
\hat n_2(\bm k,N)
+
\hat n_3(\bm k,N)
-
\hat n_0(\bm k,N)
\big)\label{Pn-R-N}
\ee
Hence
\be
e^{iP(N)\cdot x}a_1(\bm R,\bm k,N)e^{-iP(N)\cdot x} &=& a_1(\bm R,\bm k,N) e^{-ix\cdot k},\\
e^{iP(N)\cdot x}a_2(\bm R,\bm k,N)e^{-iP(N)\cdot x} &=& a_2(\bm R,\bm k,N) e^{-ix\cdot k},\\
e^{iP(N)\cdot x}a_3(\bm R,\bm k,N)e^{-iP(N)\cdot x} &=& a_3(\bm R,\bm k,N) e^{-ix\cdot k},\\
e^{iP(N)\cdot x}a_0(\bm R,\bm k,N)^{\dag}e^{-iP(N)\cdot x} &=& a_0(\bm R,\bm k,N)^{\dag} e^{-ix\cdot k},
\ee
and
\be
U(\bm 1,y,N)^{\dag}A_{a}(x,N)U(\bm 1,y,N) &=&
A_{a}(x-y,N).
\ee
Boosts and rotations are represented by means of the operator
\be
U(\Lambda,0,1) &=&\int d\tilde k{\,}d\tilde R{\,}|\bm R,\bm k\rangle\langle
\bm{\Lambda^{-1}R},\bm{\Lambda^{-1}k}|\otimes
e^{-2i \Theta(\Lambda,\bm k) L_3}.\label{U-R}
\ee
The operator
\be
L_{3}
=
J_3
=
i(a_{2}^{\dagger}a_{1}-a_{1}^{\dagger}a_{2})=
-(a_{+}^{\dagger}a_{+}-a_{-}^{\dagger}a_{-})\label{L=J}
\ee
is normally ordered. This should be contrasted with the standard-form operator (\ref{sf U}) that includes $L_1$, $L_2$,  which do not annihilate vacuum states.
The required transformation rule
\be
U(\Lambda,0,1)^{\dag}A_a(x,1)U(\Lambda,0,1)=\Lambda{_a}{^b}
A_b\big(\Lambda^{-1}x,1\big)
\ee
extends to arbitrary $N$ by (\ref{PGN}). Finally, in momentum space we find
\be
\left(
\begin{array}{c}
U(\Lambda,0,N)^{\dag}b_+(\bm R,\bm k,N)U(\Lambda,0,N)\\
U(\Lambda,0,N)^{\dag}a_+(\bm R,\bm k,N)U(\Lambda,0,N)\\
U(\Lambda,0,N)^{\dag}a_-(\bm R,\bm k,N)U(\Lambda,0,N)\\
U(\Lambda,0,N)^{\dag}b_-(\bm R,\bm k,N)U(\Lambda,0,N)
\end{array}
\right)
&=&
\left(
\begin{array}{cccc}
1 & 0 & 0 & 0\\
0 & e^{2i\Theta(\Lambda,\bm k)} & 0 & 0\\
0 & 0 & e^{-2i\Theta(\Lambda,\bm k)} & 0\\
0 & 0 & 0 & 1
\end{array}
\right)
\left(
\begin{array}{c}
b_+(\bm{\Lambda^{-1}R},\bm{\Lambda^{-1}k},N)\\
a_+(\bm{\Lambda^{-1}R},\bm{\Lambda^{-1}k},N)\\
a_-(\bm{\Lambda^{-1}R},\bm{\Lambda^{-1}k},N)\\
b_-(\bm{\Lambda^{-1}R},\bm{\Lambda^{-1}k},N)
\end{array}
\right).
\label{APM-R}
\ee
The field $A_a(x,N)$ is a direct sum of zero-mass representations: The two spin-1 unitary representations corresponding to transverse photons, and two spin-0 unitary representations corresponding to ``longitudinal" and ``timelike" ``photons". The latter two fields do not mix with the usual photons, but there is no reason to regard them as unphysical particles. The representation of CCR is reducible, but the representations of the Poincar\'e group are irreducible.

Let us stress that now there exists the following Poincar\'e invariant splitting of the four-momentum (\ref{Pn-R-N}),
\be
P_a(N)
&=&
P_a(N)_{\rm EM}+P_a(N)_{\rm S},\\
P_a(N)_{\rm EM}
&=&
{\textstyle\int} d\tilde k{\,}k_a
\big(
\hat n_1(\bm k,N)
+
\hat n_2(\bm k,N)
\big),\\
P_a(N)_{\rm S}
&=&
{\textstyle\int} d\tilde k{\,}k_a
\big(
\hat n_3(\bm k,N)
-
\hat n_0(\bm k,N)
\big),
\ee
into electromagnetic and scalar parts. This is possible only in COM representations.

\section{Correspondence with classical electrodynamics: Fields produced by classical currents in COM $N$-representations}

In COM $N$-representations Lorentz transformations do not mix transverse and timelike/longitudinal degrees of freedom, and thus one can separately consider two types of displacement operators:
\be
{\cal D}_{12}(\alpha,N)
&=&
\exp
\int d\tilde k{\,}d\tilde R\Big(
\overline{\alpha_{1}(\bm R,\bm k)} a^1(\bm R,\bm k,N)
+
\overline{\alpha_{2}(\bm R,\bm k)} a^2(\bm R,\bm k,N)
-
{\rm H.c.}
\Big)={\cal D}_{12}(\alpha/\sqrt{N},1)^{\otimes N},\\
{\cal D}_{03}(\alpha,N)
&=&
\exp
\int d\tilde k{\,}d\tilde R\Big(
\overline{\alpha_{3}(\bm R,\bm k)} a^3(\bm R,\bm k,N)
+
\overline{\alpha_{0}(\bm R,\bm k)} a^0(\bm R,\bm k,N)^{\dag}
-
{\rm H.c.}
\Big)={\cal D}_{03}(\alpha/\sqrt{N},1)^{\otimes N}.
\ee
A correspondence principle with classical electrodynamics is obtained by coherent-state averages with
$\alpha_{1}(\bm R,\bm k)=\alpha_{1}(\bm k)$,
$\alpha_{2}(\bm R,\bm k)=\alpha_{2}(\bm k)$,
$\alpha_{3}(\bm R,\bm k)=0$, $\alpha_{0}(\bm R,\bm k)=0$, and $|0,N\rangle=|0,1\rangle^{\otimes N}$,
\be
|0,1\rangle
&=&
\int d\tilde k d\tilde R\, O_0(\bm k)O_1(\bm R)|\bm R,\bm k,0,0,0,0\rangle,\label{0,1 R}\\
\int d\tilde k |O_0(\bm k)|^2
&=&
\int d\tilde R |O_1(\bm R)|^2=1.
\ee
Indeed, let $|\alpha,N\rangle={\cal D}_{12}(\alpha,N)|0,1\rangle$ be such a coherent state. The solution of the Heisenberg equation for a pointlike charge is
\be
A_a(x,J,N)
&=&
A_a(x,N)
+
q
\int_{0}^{\sqrt{x^2}}ds \frac{dX_a(s)}{ds}
\hat D(x-X(s),N)\\
\ee
where $q$ is the {\it bare\/} charge occurring in the interaction Hamiltonian.
The coherent-state average reads
\be
\langle\alpha,N|A_a(x,J,N)|\alpha,N\rangle
&=&
\langle\alpha,N|A_a(x,N)|\alpha,N\rangle
+
q
\int_{0}^{\sqrt{x^2}}ds \frac{dX_a(s)}{ds}
\langle\alpha,N|\hat D(x-X(s),N)|\alpha,N\rangle\label{classical1}\\
&=&
\langle\alpha,1|A_a(x,1)|\alpha,1\rangle
+
q
\int_{0}^{\sqrt{x^2}}ds \frac{dX_a(s)}{ds}
\langle 0,1|\hat D(x-X(s),1)|0,1\rangle.\label{classical2}
\ee
The equivalence between (\ref{classical1}) and (\ref{classical2}) follows from the assumed form of the vacuum state and the fact that displacements operators commute with $I(\bm R,\bm k,N)$. This correspondence principle is thus insensitive to $N$. We will later see that there exists another correspondence principle, mathematically expressed by the weak limit $N\to\infty$, linking $N$-representations with {\it regularized\/} forms of irreducible representations.

Let us first have a look at the free-field part.
Employing (\ref{A-second}), denoting $Z_0(\bm k)=|O_0(\bm k)|^2$,
$Z_1(\bm k)=|O_1(\bm k)|^2$, we find
\be
\langle\alpha,1|A_a(x,1)|\alpha,1\rangle
&=& i\int d\tilde k{\,}Z_0(\bm k)\left(
  \langle g_{a}{^1}(\bm k)\rangle \alpha_1(\bm k)
+ \langle g_{a}{^2}(\bm k)\rangle \alpha_2(\bm k)
\right) e^{-ik\cdot x}+{\rm c.c},\nonumber\\
\langle g_{a}{^j}(\bm k)\rangle
&=& \int d\tilde R\,Z_1(\bm R) g_{a}{^j}(\bm R,\bm k).
\ee
This is a classical four-potential with momentum-space amplitudes $Z_0(\bm k)\alpha_j(\bm k)$ and transverse linear polarization vectors
$\langle g_{a}{^j}(\bm k)\rangle$. Switching to the null tetrad we obtain
\be
\langle\alpha,1|A_a(x,1)|\alpha,1\rangle
&=& -i\int d\tilde k{\,}Z_0(\bm k)\Big(
  \langle \omega_{A}(\bm k)\rangle \bar\pi_{A'}(\bm k)\alpha_+(\bm k)
+ \pi_{A}(\bm k) \langle\bar\omega_{A'}(\bm k)\rangle \alpha_-(\bm k)
\Big) e^{-ik\cdot x}+{\rm c.c}.
\ee
Since
\be
\langle \omega_{A}(\bm k)\rangle\pi^{A}(\bm k)
&=&
\int d\tilde R\, Z_1(\bm R)\omega_{A}(\bm R,\bm k)\pi^{A}(\bm k)
=
\int d\tilde R\, Z_1(\bm R)=1
\ee
the pair $\langle \omega_{A}(\bm k)\rangle$, $\pi_{A}(\bm k)$ is again a spin-frame. Such a free-field potential satisfies the Lorenz gauge
\be
\partial^a \langle\alpha,1|A_a(x,1)|\alpha,1\rangle=\partial^a \langle\alpha,N|A_a(x,N)|\alpha,N\rangle=0.
\ee
The corresponding free-field tensor
\be
F_{ab}(x)
&=&
\partial_a \langle\alpha,1|A_b(x,1)|\alpha,1\rangle-\partial_b\langle\alpha,1|A_a(x,1)|\alpha,1\rangle\\
&=&
\ve_{AB}
\int d\tilde k{\,}Z_0(\bm k)
\bar\pi_{A'}(\bm k)\bar\pi_{B'}(\bm k)
\Big(
\alpha_+(\bm k)e^{-ik\cdot x}
+
\overline{\alpha_-(\bm k)}e^{ik\cdot x}
\Big)
\nonumber\\
&\pp=&
+
\ve_{A'B'}
\int d\tilde k{\,}Z_0(\bm k)
\pi_{A}(\bm k)\pi_{B}(\bm k)
\Big(
\alpha_-(\bm k)e^{-ik\cdot x}
+
\overline{\alpha_+(\bm k)} e^{ik\cdot x}
\Big)\label{F}
\ee
has the standard form typical of electromagnetic free fields with momentum-space amplitudes
$Z_0(\bm k)\alpha_\pm(\bm k)$.

Now, let us turn to the source term in (\ref{classical2}) and assume, similarly to Section \ref{Fields-irr}, that $X_a(s)=u_as$, where $u_a$ is a constant four-velocity. The solution of the Heisenberg equation then reads explicitly
\be
A_a(x,J,N)
&=&
A_a(x,N)
+
2qu_a
\int d\tilde k\,d\tilde R\,
I(\bm R,\bm k,N)\frac{\cos[k\cdot (x-u\sqrt{x^2})]-\cos[k\cdot x]}{k\cdot u}.
\ee
The source term, when averaged in $|\alpha,N\rangle$, reduces to
\be
2qu_a
\int d\tilde k\,
Z_0(\bm k)\frac{\cos[k\cdot (x-u\sqrt{x^2})]-\cos[k\cdot x]}{k\cdot u}.
\ee
Let $Z=\max_{\bm k}\{Z_0(\bm k)\}$. The form of $U(\Lambda,y,N)$ for COM $N$-representations implies that  $Z$ is Poincar\'e invariant since
$\max_{\bm k}\{Z_0(\bm k)\}=\max_{\bm k}\{Z_0(\bm {\Lambda^{-1}k})\}$. Denoting $\chi(\bm k)=Z_0(\bm k)/Z$, $q_{\rm ren}=Z^{1/2}q$, $A_{\rm ren}^a(x)=Z^{-1/2}A^a(x)$, $A_a{_{\rm ren}}(x,J)=Z^{-1/2}A_a(x,J)$,
and $\alpha_{\pm\rm ren}(\bm k)=Z^{1/2}\alpha_\pm(\bm k)$, we get
\be
\langle\alpha,N|A_{a\,\rm ren}(x,J,N)|\alpha,N\rangle
&=& -i\int d\tilde k{\,}\chi(\bm k)\Big(
  \langle \omega_{A}(\bm k)\rangle \bar\pi_{A'}(\bm k)\alpha_{+\rm ren}(\bm k)
+ \pi_{A}(\bm k) \langle\bar\omega_{A'}(\bm k)\rangle \alpha_{-\rm ren}(\bm k)
\Big) e^{-ik\cdot x}+{\rm c.c}.
\\
&\pp=&
+2q_{\rm ren}u_a
\int d\tilde k\,
\chi(\bm k)\frac{\cos[k\cdot (x-u\sqrt{x^2})]-\cos[k\cdot x]}{k\cdot u}.
\ee
Let us note that replacing  $\chi(\bm k)$ by 1 we obtain {\it exactly\/} the formula occurring for irreducible representations with $Z\neq 1$.
However, since $\int d\tilde k\,Z_0(\bm k)=\int d\tilde k\,|O_0(\bm k)|^2=1$ the function $0\leq\chi(\bm k)\leq 1$ must vanish for $\bm k\to\infty$. In consequence, the fact that $O_0(\bm k)\in L^2(d\tilde k)$ implies that the theory gets {\it automatically\/} ultraviolet-regularized. An infrared regularization, on the other hand, is equivalent to the Poincar\'e invariant boundary condition $O_0(\bm 0)=0$. The invariant $Z$ plays a role of the renormalization constant $Z_3$.

Let us make here a remark that the average electromagnetic fields we have found by quantizing in the COM $N$-representation do not differ from the formulas we would have found if we had performed the computation in the standard-form $N$-representation. The differences are at the level of relativistic transformations of fields (in momentum representation), and relativistic invariance of the vacuum subspaces of states (the latter holds only for COM representations). The representation introduced in \cite{II} (standard-form $N$-representations but involving only the two transverse polarizations) led to the correct zero-mass spin-1 unitary transformations of momentum-space transverse polarizations and invariance of the vacuum subspace, but in position representation the four-potential was not a four-vector field. The COM representations are free of all these difficulties, and yet are in many respects practically indistinguishable from the standard-form $N$-representations.

\section{Static pointlike charge in spherically symmetric vacuum}

Let us now consider the special case of the rest frame of a single bare charge $q$, i.e. the one with $(u_0,u_1,u_2,u_3)=(1,0,0,0)$, and the vacuum state $|0,N\rangle$ whose probability density in the $\bm k$-space is in this reference frame spherically symmetric. In order to simplify calculations we assume that $Z_0(\bm k)=0$ for $|\bm k|<k_1$ and $|\bm k|>k_2$, and $Z_0(\bm k)=Z=$ const for $k_1\leq|\bm k|\leq k_2$, implying
$q_{\rm ren} = Z^{1/2}q=\frac{2\sqrt{2}\pi }{\sqrt{k_2^2-k_1^2}}q$.

We are interested only in the source term, and this is unaffected by the choice of $\alpha$ in $|\alpha,N\rangle$, so let us take $\alpha=0$. Then
the only non-vanishing component of $\langle\alpha,N|A_{a\,\rm ren}(x,J,N)|\alpha,N\rangle$ is
\be
{}&{}&\langle 0,N|A_{0\,\rm ren}(\sqrt{\tau^2+\bm x^2},\bm x,J,N)|0,N\rangle\nonumber\\
&\pp=&\pp=
=
\frac{q_{\rm ren}}{(2\pi)^2|\bm x|}
\Big(
{\rm si\,}k_2(x_0-\tau+|\bm x|)-{\rm si\,}k_1(x_0-\tau+|\bm x|)
+
{\rm si\,}k_2(|\bm x|-x_0+\tau)-{\rm si\,}k_1(|\bm x|-x_0+\tau)
\nonumber\\
&\pp=&
\pp{===\frac{q_{\rm ph}}{(2\pi)^2|\bm x|}\Big(}
-
{\rm si\,}k_2(x_0+|\bm x|)
+
{\rm si\,}k_1(x_0+|\bm x|)
+
{\rm si\,}k_2(x_0-|\bm x|)
-
{\rm si\,}k_1(x_0-|\bm x|)
\Big)\label{Coulomb-N}
\ee
for $\tau\geq 0$. The solution, as opposed to the irreducible case, is continuous in $\tau$ and has no singularity at $\bm x=0$:
\be
\langle 0,N|A_{0\,\rm ren}(\tau,\bm 0,J,N)|0,N\rangle
&=&
\frac{q_{\rm ren}}{2\pi^2}
\Big(
k_2-k_1+\frac{\sin k_1\tau}{\tau}-\frac{\sin k_2\tau}{\tau}
\Big)\label{Coulomb-N0}.
\ee
Employing
\be
\lim_{\tau\to\infty}{\rm si\,}k(\pm\sqrt{\tau^2+x^2}\mp\tau+x) &=&{\rm si\,}k x,\nonumber\\
\lim_{\tau\to\infty}{\rm si\,}k(\sqrt{\tau^2+x^2}+x) &=&\pi/2\nonumber
\ee
we get the asymptotic form
\be
\lim_{\tau\to\infty}\langle 0,N|A_{0\,\rm ren}(\sqrt{\tau^2+\bm x^2},\bm x,J,N)|0,N\rangle
&=&
\frac{q_{\rm ren}}{2\pi^2}
\frac{{\rm si\,}k_2|\bm x|-{\rm si\,}k_1|\bm x|}{|\bm x|}.\label{Coulomb-N-infty}
\ee
If $k_1=0$ then spacelike asymptotics is exactly Coulombian
\be
\frac{q_{\rm ren}}{2\pi^2}
\frac{1}{|\bm x|}{\rm si\,}k_2|\bm x|
\to
\frac{q_{\rm ren}}{2\pi^2}
\frac{1}{|\bm x|}\frac{\pi}{2}
=
\frac{q_{\rm ren}}{4\pi |\bm x|}
\ee
with $|\bm x|\to \infty$. If $k_1\neq 0$, i.e. when $Z_0(\bm 0)=0$, the potential decays faster than $1/|\bm x|$. Fig.~1 compares (\ref{Coulomb-N-infty}) with the Coulomb law for $k_2=10^4$ (in arbitrary units) and various values of $k_1$. Deviations from the Coulomb law at large distances allow, at least in principle, to set an upper bound on $k_1$.

\begin{figure}\label{Fig1}
\includegraphics[width=12 cm]{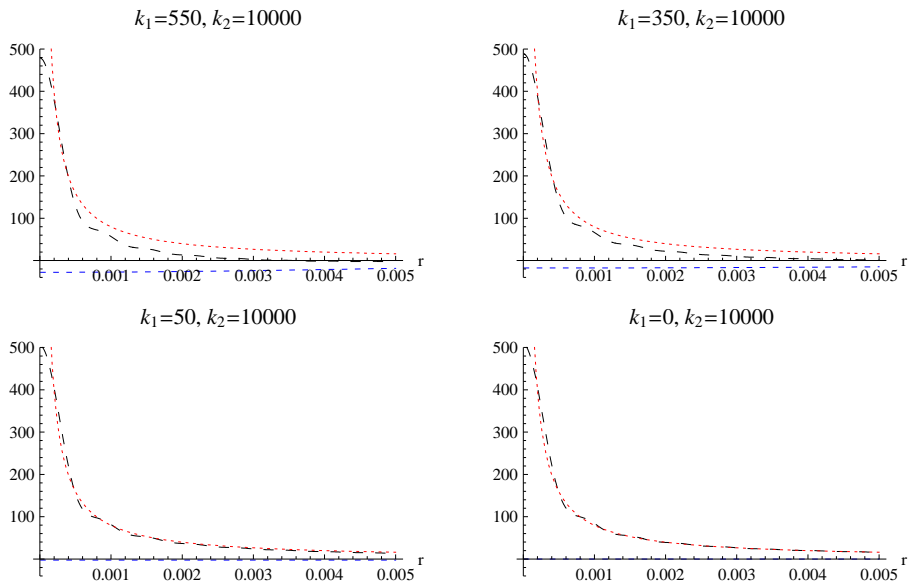}
\caption{(Color online) The potentials as functions of $r=|\bm x|$. The Coulomb field is red-dotted. Black medium-dashed is the $\tau\to\infty$ limit of the potential following from the COM $N$-representation. The blue short-dashed curve shows for comparison the part $-\frac{q_{\rm ren}}{2\pi^2}
\frac{{\rm si\,}k_1r}{r}$. Units are chosen arbitrarily and $q_{\rm ren}=1$.}
\end{figure}

Let us stress it again that modifications of the Coulomb law are here a consequence of the structure of the quantum vacuum in reducible $N$-representations,
and not of modifications of the current $J_a(x)$ entering the Hamiltonian (\ref{H1 tau}). The current in this example is the usual classical, pointlike, static one. The photon field is also exactly massless.

So how is it possible that without modification of the current or introducing a photon mass we have found a modification of the Coulomb field?

The answer is simple: We have not used the Maxwell equations at any stage but worked exclusively with the Heisenberg picture. The link with Maxwell's equations can be established by first computing the average Maxwell field
\be
F_{ab}(x)
&=&
\partial_a \langle\alpha,N|A_{b\,\rm ren}(x,J,N)|\alpha,N\rangle-\partial_b\langle\alpha,N|A_{a\,\rm ren}(x,J,N)|\alpha,N\rangle
\ee
and then defining the effective conserved current
\be
J_{b\,\rm eff}(x)
&=&
\partial^a
F_{ab}(x).
\ee
We will not give here the explicit form of $J_{b\,\rm eff}(x)$ as the formula is not very illuminating. But in the limit $\tau\to\infty$ the effective charge density takes a simple form,
\be
\rho_{\rm eff}(\bm x)=q_{\rm ren}
\frac{\sin(k_2 |\bm x|)- k_2 |\bm x| \cos (k_2 |\bm x|)}{2 \pi^2 |\bm x|^3}
-
q_{\rm ren}
\frac{\sin(k_1 |\bm x|) -k_1 |\bm x| \cos (k_1 |\bm x|)}{2 \pi^2 |\bm x|^3}
.\label{rho eff}
\ee
Actually, this type of charge density is even too simple since the integral
\be
Q&=&\int \rho_{\rm eff}(\bm x)d^3x
=
q_{\rm ren}\lim_{r\to\infty}
\frac{2}{\pi}
\big(
\sin(k_1 r) - \sin(k_2 r)  + {\rm si\,}(k_2 r)- {\rm si\,}(k_1 r)
\big)\label{qeff}
\ee
is ill defined due to the oscillating first two terms. Replacing the discontinuities in $Z_0(\bm k)$ by continuous approximations to the step function, which is of course more realistic, we obtain $\rho_{\rm eff}(\bm x)$ that approximates (\ref{rho eff}) with arbitrary accuracy but the integral at the left-hand side of (\ref{qeff}) becomes well defined. We then find two cases: Either $Z_0(\bm 0)=0$ and $Q=0$, or $Z_0(\bm 0)>0$ and $Q=q_{\rm ren}$. The same conclusion was found in \cite{CN} where the instant form dynamics was used.

Now which case is more physical: Total charge $Q=q_{\rm ren}$ or $Q=0$? We believe there are many reasons to impose $Z_0(\bm 0)=0$ and, accordingly, $Q=0$. First of all, waves of infinite wavelength and zero frequency are unphysical. When we arrive at photon statistics it will turn out that $Z_0(\bm 0)=0$ is needed to avoid infrared divergences. Moreover, the representations of the Poincar\'e group with $k^a=0$ have to be treated separately from those with  $k^2=0$, $k^a\neq 0$ (are induced from different little groups and $k^a=0$ and $k^a\neq 0$ do not belong to the same homogeneous space), and so on. So we have to understand $Q=0$.

Plotting $\lim_{\tau\to\infty}\langle 0,N|A_{0\,\rm ren}(\sqrt{\tau^2+\bm x^2},\bm x,J,N)|0,N\rangle$ and $\rho_{\rm eff}(\bm x)$ for various values of $k_1$ and $k_2$ we observe that the cases $k_1=0$ and $k_1\approx 0$ are locally practically indistinguishable. Fig.~2 illustrates the modifications of $\rho_{\rm eff}(\bm x)$ when we change $k_1$. For $k_2=10^3$ there is no visible difference between $k_1=0$ and $k_1=150$, in spite of the fact that globally the two charge distributions are different. Let us also note that the plots are made under the assumption $q_{\rm ren}=1$, but the charge density involves both signs of charge.

\begin{figure}\label{Fig2}
\includegraphics[width=12 cm]{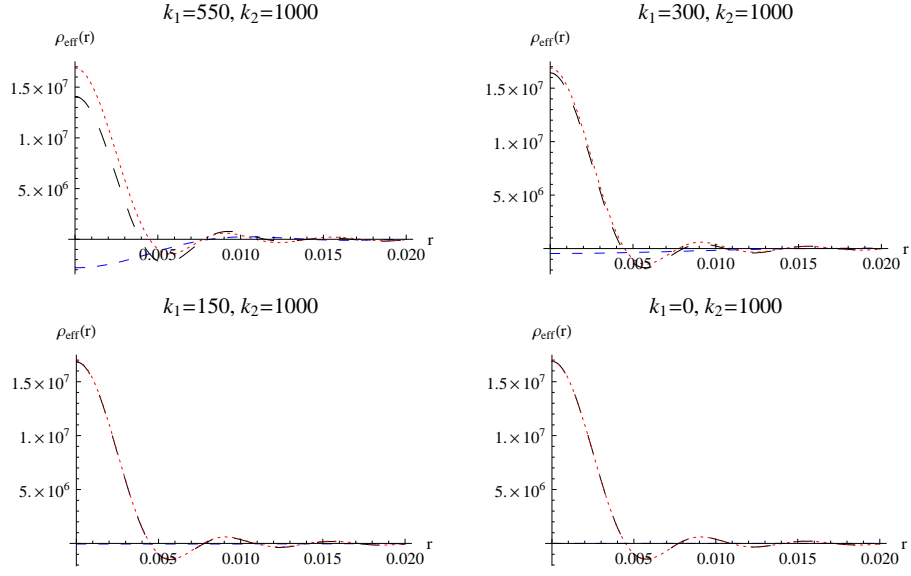}
\caption{(Color online) The effective charge density (black, long-dashed) is a sum of two terms that generate total charges of opposite sign: One density (short-dashed, blue) that becomes the more uniform as a function of $r=|\bm x|$ the smaller $k_1$, and the other (dotted, red) that becomes the more localized the greater $k_2$. The long-dashed and dotted curves become practically indistinguishable even for a relatively large $k_1$. However, for $k_1=0$ one of the two densities vanishes, and thus the total charge $Q=q_{\rm ren}$ is nonzero. Units on the plots are chosen arbitrarily.}
\end{figure}

\section{Photon statistics in COM $N$-representations}

The analysis given in the previous section shows that pointlike charges may generate effective classical fields that look like being produced by extended sources. What is interesting the effect crucially depends on the choice of representation of CCR and its corresponding vacuum state, and thus is fundamentally {\it quantum\/}. But to relate the choice of $N$-representation with quantum optical experiments we need to discuss photon statistics. More precisely, we are interested only in the statistics of transverse photons. The ``longitudinal" and ``timelike" scalar particles are assumed to be undetectable in optical measurements.

To begin with, let us recall that an $n$-photon state is the one that contains $n$ transverse-polarization excitations distributed, in $N$-representations, over all the $N$ oscillators. Such states are spanned by vectors of the form (\ref{n-fot}) (appropriately modified in COM representations by the presence of $\bm R_1$$\dots$$\bm R_N$). For $N=1$ the probability of finding $n_+^{(1)}$ right-handed and $n_-^{(1)}$ left-handed excitations is given by an average of the projector
\be
\Pi(n_+^{(1)},n_-^{(1)})
&=&
I^{(1)}\otimes |n_+^{(1)}\rangle\langle n_+^{(1)}|\otimes |n_-^{(1)}\rangle\langle n_-^{(1)}| \otimes I_3^{(1)}\otimes I_0^{(1)}
\ee
where $I^{(1)}=\int d\tilde k d\tilde R|\bm R,\bm k\rangle\langle\bm R,\bm k|$. For arbitrary $N$ the corresponding projector is
\be
\Pi(n_+,n_-)
=
\sum_{n_+^{(1)},\dots,n_-^{(N)}}
\Pi(n_+^{(1)},n_-^{(1)})\otimes\dots \otimes \Pi(n_+^{(N)},n_-^{(N)})\label{Proj}
\ee
where we sum over all combinations of numbers of excitations that satisfy $\sum_{j=1}^Nn_\pm^{(j)}=n_\pm$.

An example of an $n$-photon state is a state that arises if one acts on a vacuum $n$ times with creation operators. So the situation is, in this respect, completely analogous to what we are accustomed to from standard quantum field theories. However, it is very important to bear in mind that in reducible representations the set of states containing $n$ excitations is larger than the set of states obtained by acting on vacuum with creation operators. To understand why it is so it is enough to act on an $n$-photon state with any element of the center of the CCR algebra. In irreducible representations all central elements are proportional to identities, but in reducible cases these elements are nontrivial and yet do not change the number of excitations.

The operator $U_1(\tau)=U_1(\tau,0)$ given by (\ref{U_1}) can be written as
\be
U_1(\tau) &=& e^{i\varphi(\tau)}V_{03}(\tau)V_{12}(\tau)
\ee
where $\varphi(\tau)$ is in the center of CCR, and $V_{\bf ab}(\tau)$ are exponents containing only the polarizations corresponding to the concrete choice of the indices $\bf a$, $\bf b$. The projector (\ref{Proj}) commutes with $e^{i\varphi(\tau)}V_{03}(\tau)$.
Let us denote by $A^\perp_{a}(x,N)$ the transverse part of $A_{a}(x,N)$,
\be
A^\perp_{a}(x,N)
&=& -i\int d\tilde k{\,} d\tilde R\Big(
  \omega_A(\bm R,\bm k)\bar\pi_{A'}(\bm k) a_+(\bm R,\bm k,N)
+ \pi_{A}(\bm k)\bar\omega_{A'}(\bm R,\bm k) a_-(\bm R,\bm k,N)
\Big) e^{-ik\cdot x}+{\rm H.c.}
\label{B-second}\\
&=& -i\int d\tilde k{\,} d\tilde R\Big(
  x_a(\bm R,\bm k)a_1(\bm R,\bm k,N)
+ y_a(\bm R,\bm k) a_2(\bm R,\bm k,N)
\Big) e^{-ik\cdot x}+{\rm H.c.},
\label{B-second'}\\
V_{12}(\tau,N)
&=&
\exp\Big(-i{\int}_{\Sigma_{\tau_0}}^{\Sigma_\tau} d^4xJ^a(x)A^\perp_{a}(x,N)
\Big).
\ee
If at $\tau=0$ the field is in a vacuum state then the probability of finding at an arbitrary $\tau$ $n_+$ and $n_-$ photons of appropriate circular polarization type is
\be
p(n_+,n_-,\tau,N)
&=&
\langle 0,N|V_{12}(\tau,N)^{\dag}\Pi(n_+,n_-)V_{12}(\tau,N)|0,N\rangle.
\label{prob}
\ee
The fact that
$a_{1}^{\dagger}a_{1}+a_{2}^{\dagger}a_{2}=a_{+}^{\dagger}a_{+}+a_{-}^{\dagger}a_{-}$
implies $n_1+n_2=n_++n_-$.
Let us for simplicity write explicitly only the formula for the total number of photons, $p(n,\tau,N)=\sum_{n_++n_-=n}p(n_+,n_-,\tau,N)$. Standard computations involving Baker-Hausdorff formula lead to
\be
p(n,\tau,N)
&=&
\frac{1}{n!}\frac{d^n}{d\lambda^n}\langle 0,N|e^{\lambda{\int}_{\Sigma_{\tau_0}}^{\Sigma_\tau} d^4xd^4yJ^a(x)J^b(y)
\int d\tilde k{\,} d\tilde R
I(\bm R,\bm k,N)
\big(x_a(\bm R,\bm k)x_b(\bm R,\bm k)
+ y_a(\bm R,\bm k)y_b(\bm R,\bm k)\big)
e^{ik\cdot (x-y)}}|0,N\rangle\Big|_{\lambda=-1}\nonumber\\
&=&
\frac{1}{n!}\frac{d^n}{d\lambda^n}C(\lambda,\tau,N)\Big|_{\lambda=-1}\nonumber
\ee
where
\be
C(\lambda,\tau,N)
&=&
\Big(\int d\tilde k{\,} d\tilde R
Z_0(\bm k)Z_1(\bm R)e^{\frac{\lambda}{N}{\int}_{\Sigma_{\tau_0}}^{\Sigma_\tau} d^4xd^4yJ^a(x)J^b(y)
\big(x_a(\bm R,\bm k)x_b(\bm R,\bm k)
+ y_a(\bm R,\bm k)y_b(\bm R,\bm k)\big)
e^{ik\cdot (x-y)}}\Big)^N. \label{C N}
\ee
The limiting case $N\to\infty$
\be
C(\lambda,\tau,\infty)
&=&
\exp \Bigg(\lambda\int d\tilde k{\,} d\tilde R
Z_0(\bm k)Z_1(\bm R){\int}_{\Sigma_{\tau_0}}^{\Sigma_\tau} d^4xd^4yJ^a(x)J^b(y)
\big(x_a(\bm R,\bm k)x_b(\bm R,\bm k)
+ y_a(\bm R,\bm k)y_b(\bm R,\bm k)\big)
e^{ik\cdot (x-y)}\Bigg)\nonumber\\\label{C inf}
\ee
implies the Poisson statistics. The case of finite $N$ is, as noticed in \cite{CN}, the R\'enyi $\alpha$-statistics with $\alpha=1-1/N$. The limit $\alpha\to 1$ is known in information theory as the Shannon limit. Indeed, one recognizes in $C(\lambda,\tau,N)$ the Kolmogorov-Nagumo average \cite{CN02} in the form used by R\'enyi in his derivation of $\alpha$-entropies \cite{Renyi}. The Shannon limit can be expressed also directly in terms of the renormalized current $J_{\rm ren}^a(x)=Z^{1/2}J^a(x)$ and $\chi(\bm k)=Z_0(\bm k)/Z$ (cf. Section~X). First of all, denoting
\be
\langle x_a(\bm k)x_b(\bm k)\rangle
+
\langle y_a(\bm k)y_b(\bm k)\rangle
&=&
\int d\tilde R\,Z_1(\bm R)\Big(x_a(\bm R,\bm k)x_b(\bm R,\bm k)
+ y_a(\bm R,\bm k)y_b(\bm R,\bm k)\Big),
\ee
we get the limiting case
\be
C(\lambda,\tau,\infty)
&=&
\exp \Bigg(\lambda\int d\tilde k{\,}
\chi(\bm k){\int}_{\Sigma_{\tau_0}}^{\Sigma_\tau} d^4xd^4yJ_{\rm ren}^a(x)J_{\rm ren}^b(y)
e^{ik\cdot (x-y)}
\Big(
\langle x_a(\bm k)x_b(\bm k)\rangle
+
\langle y_a(\bm k)y_b(\bm k)\rangle
\Big)
\Bigg)
\ee
which is the {\it regularized\/} form of the expression we would have obtained for irreducible representations. The regularization is here automatic and does not have to be justified by {\it ad hoc\/} arguments on unobservability of ``soft photons", of course provided $\chi(\bm 0)=0$. 

\section{Photon statistics for a pointlike source}

Let us return to the pointlike current (\ref{J point}) and consider $s\mapsto X^a(s)=u^a s$ where $u^a$ is a constant four-velocity. Then
\be
{\int}_{\Sigma_{\tau_0}}^{\Sigma_\tau} d^4xd^4yJ_{\rm ren}^a(x)J_{\rm ren}^b(y)e^{ik\cdot (x-y)}
&=&
q_{\rm ren}^2
\int_{0}^{\tau}ds \frac{dX^a(s)}{ds}e^{ik\cdot X(s)}\int_{0}^{\tau}ds'
\frac{dX^b(s')}{ds'}e^{-ik\cdot X(s')}\\
&=&
q_{\rm ren}^2 u^au^b
\frac{\sin^2 (k\cdot u\tau/2)}{(k\cdot u/2)^2}
\ee
In particular, the $N\to\infty$ vacuum-to-vacuum probability reads
\be
C(-1,\tau,\infty)
&=&
\exp \Bigg(-q_{\rm ren}^2 \int d\tilde k{\,}
\chi(\bm k)
\frac{\sin^2 (k\cdot u\tau/2)}{(k\cdot u/2)^2}
u^au^b
\Big(
\langle x_a(\bm k)x_b(\bm k)\rangle
+
\langle y_a(\bm k)y_b(\bm k)\rangle
\Big)
\Bigg).\label{C -1}
\ee
For $\tau>0$ this probability is smaller than 1, and thus a uniformly moving pointlike classical charge emits photons.
For large times, $\tau\to \infty$,
\be
C(-1,\infty,\infty)
&=&
\exp \Bigg(-2q_{\rm ren}^2 \int d\tilde k{\,}
\chi(\bm k)
\frac{\langle [u\cdot x(\bm k)]^2\rangle
+
\langle [u\cdot y(\bm k)]^2\rangle}{(k\cdot u)^2}
\Bigg).
\ee
Now consider the simplest case of an accelerated charge. The world-line
\be
X^a(s)
&=&
\left\{
\begin{array}{lcl}
u^a s &{\rm for} &0\leq s <\tau_1\\
u^a \tau_1+ v^a(s-\tau_1) &{\rm for} &\tau_1\leq s
\end{array}
\right.
\ee
describing the change $u^a\to v^a$ of four-velocity at $\tau=\tau_1$, implies for times $\tau>\tau_1$
\be
\int_{0}^{\tau}ds \frac{dX^a(s)}{ds}e^{ik\cdot X(s)}
&=&
e^{ik\cdot u\tau_1}
\Big(
u^a\frac{1-e^{-ik\cdot u\tau_1}}{ik\cdot u}
-
v^a
\frac{1-e^{ik\cdot v(\tau-\tau_1)}}{ik\cdot v}
\Big).
\ee
Restricting the analysis only to the Shannon limit $N\to\infty$ we find
\be
C(-1,\tau,\infty)
&=&
\exp \Bigg(-2q_{\rm ren}^2\int d\tilde k{\,}
\chi(\bm k)\Big(
\langle x_a(\bm k)x_b(\bm k)\rangle
+
\langle y_a(\bm k)y_b(\bm k)\rangle
\Big)
\nonumber\\
&\pp=&\pp{\exp \Bigg(-2q_{\rm ren}^2\int }
\times\Bigg[
\frac{u^au^b}{(k\cdot u)^2}
\Big(1-\cos(k\cdot u\tau_1)\Big)
+
\frac{v^av^b}{(k\cdot v)^2}
\Big(1-\cos\big(k\cdot v(\tau-\tau_1)\big)\Big)
\nonumber\\
&\pp=&\pp{\exp \Bigg(-2q_{\rm ren}^2\int d\tilde k{\,}}
-
\frac{u^av^b}{(k\cdot u)(k\cdot v)}\Big(1-\cos\big(k\cdot u\tau_1\big)\Big)
\Big(1-\cos\big(k\cdot v(\tau-\tau_1)\big)\Big)
\nonumber\\
&\pp=&\pp{\exp \Bigg(-2q_{\rm ren}^2\int d\tilde k{\,}}
+\frac{u^av^b}{(k\cdot u)(k\cdot v)}\sin(k\cdot u\tau_1)\sin\big(k\cdot v(\tau-\tau_1)\big)
\Bigg]
\Bigg).\label{tau_1 tau}
\ee
A comparison with standard quantum optics should take into account that $\tau$ and $\tau_1$ describe the dynamics since the origin of the Universe.
Thus $\Delta\tau=\tau-\tau_1$ can be of the order of time scales available in experiments, whereas $\tau_1$ is of the order of the age of the Universe. The finite-time dynamics we encounter in present-day experiments is the limit $\tau_1\to\infty$ of (\ref{tau_1 tau}), with fixed $\Delta\tau$:
\be
\lim_{\tau_1\to\infty}C(-1,\tau_1+\Delta\tau,\infty)
&=&
\exp \Bigg(-2q_{\rm ren}^2\int d\tilde k{\,}
\chi(\bm k)\Big(
\langle x_a(\bm k)x_b(\bm k)\rangle
+
\langle y_a(\bm k)y_b(\bm k)\rangle
\Big)\label{infty tau}
\\
&\pp=&\pp{\exp}
\times\Bigg[
\frac{u^au^b}{(k\cdot u)^2}
+
\frac{v^av^b}{(k\cdot v)^2}
\Big(1-\cos(k\cdot v \Delta\tau)\Big)
-
\frac{u^av^b}{(k\cdot u)(k\cdot v)}
\Big(1-\cos(k\cdot v\Delta\tau)\Big)
\Bigg]
\Bigg).\nonumber
\ee
The prediction corresponding to the $S$-matrix vacuum-to-vacuum probability is thus
\be
{}&{}&
\lim_{\Delta\tau\to\infty}\lim_{\tau_1\to\infty}C(-1,\tau_1+\Delta\tau,\infty)
\nonumber\\
&\pp=&\pp{===}
=
\exp \Bigg(-2q_{\rm ren}^2\int d\tilde k{\,}
\chi(\bm k)\Big(
\langle x_a(\bm k)x_b(\bm k)\rangle
+
\langle y_a(\bm k)y_b(\bm k)\rangle
\Big)
\Bigg[
\frac{u^au^b}{(k\cdot u)^2}
+
\frac{v^av^b}{(k\cdot v)^2}
-
\frac{u^av^b}{(k\cdot u)(k\cdot v)}
\Bigg]
\Bigg).\label{S}
\ee
Replacing in (\ref{S}) $q_{\rm ren}^2$ by $-\lambda q_{\rm ren}^2$ we obtain a generating function of Poisson probability distribution.
The average number of photons associated with this Poisson distribution is
\be
\bar n(u,v)
&=&
q_{\rm ren}^2\int d\tilde k{\,}
\chi(\bm k)\Big(
\langle x_a(\bm k)x_b(\bm k)\rangle
+
\langle y_a(\bm k)y_b(\bm k)\rangle
\Big)
\Bigg[
\frac{u^au^b}{(k\cdot u)^2}
+
\frac{v^av^b}{(k\cdot v)^2}
\Bigg]\nonumber\\
&+&
q_{\rm ren}^2\int d\tilde k{\,}
\chi(\bm k)\Big(
\langle x_a(\bm k)x_b(\bm k)\rangle
+
\langle y_a(\bm k)y_b(\bm k)\rangle
\Big)
\Bigg[
\frac{u^a}{k\cdot u}
-
\frac{v^a}{k\cdot v}
\Bigg]
\Bigg[
\frac{u^b}{k\cdot u}
-
\frac{v^b}{k\cdot v}
\Bigg]
\ee
Let us note that the second term is exactly the {\it regularized\/} and {\it renormalized\/} Bremsstrahlung known from standard quantum radiation theory (cf. Eq.~(1-211) in \cite{IZ}). The first term contains contributions to the average number of photons from the inertial parts of the trajectory. The latter contribution is absent in more standard approaches and survives even if $u^a=v^a$; our formulation allows to treat the Coulomb part of the potential in a fully quantum way, with no need of separating it from radiation.
Finally, the cut-off function $\chi(\bm k)=|O_0(\bm k)|^2/Z$, $Z=\max_{\bm k}\{|O_0(\bm k)|^2\}$, appears here as a consequence of field quantization, does not have to be justified by {\it ad hoc\/} arguments, and should not be removed from the final result. We will return to the cutoff problem in the final Section.

We do not explicitly discuss the R\'enyi statistics occurring for finite $N$, but its derivation is straightforward.

\section{Covariance and invariance of the formalism}

The Milne unverse itself is covariant under four-translations only in the restricted sense of passive changes of coordinates in the background Minkowski space (i.e. one has to shift also the origin $x^a=0$). The internal symmetry group of the Milne universe thus consists of SL(2,C) transformations and the shifts $\tau\to\tau+\Delta\tau$. The latter are given by the dynamics, so we have to restrict the analysis to the SL(2,C) part.

The representation of the Poincar\'e group was constructed for free fields. These fields, in the interaction picture, are the basic objects that define interaction at $\tau=\tau_0$. If one analogously constructs a fermionic representation $U_F(\Lambda,y,N_F)$ of the group \cite{III}, one arrives at a quantum current satisfying
\be
U_F(\Lambda,y,N_F)^{\dag}J_a(x,N_F)U_F(\Lambda,y,N_F)=\Lambda{_a}{^b}J_b\big(\Lambda^{-1}(x-y),N_F\big),
\ee
where $N_F$ is independent of the bosonic parameter $N_B=N$ we employ in the present paper. Therefore, the formalism we have introduced becomes internally consistent only if we assume that the Poincar\'e transformation  $U(\Lambda,y,N)$ of electromagnetic fields is accompanied by
\be
J_a(x)\mapsto J'_a(x)=\Lambda{_a}{^b}J_b\big(\Lambda^{-1}(x-y)\big),
\ee
plus the required transformation of the region of integration in the source term.
Assuming this, let us investigate which quantities we have discussed are invariant, and which are only covariant. We concentrate only on the SL(2,C) part.

\subsection{Covariance of $A_a(x,J,N)$ in COM $N$-representations}

In both standard-form and COM form of $N$-representations one finds
\be
U(\Lambda,y,N)^{\dag}\hat D(x,N)U(\Lambda,y,N)=\hat D(\Lambda^{-1}x,N).
\ee
Accordingly,
\be
U(\Lambda,0,N)^{\dag}A_a(x,J',N)U(\Lambda,0,N)
&=&
\Lambda{_a}{^b}A_b(\Lambda^{-1}x,J,N).
\ee
\subsection{Invariance of the space of vacuum states}

Acting with (\ref{U-R}) on (\ref{0,1 R}) we find
\be
U(\Lambda,0,1)|0,1\rangle
&=&
\int d\tilde k d\tilde R\, O_0(\bm{\Lambda^{-1} k})O_1(\bm{\Lambda^{-1} R})|\bm R,\bm k,0,0,0,0\rangle.
\ee
A transformed vacuum state is again a vacuum state, but the probability of finding $\bm k$ is modified by the Doppler effect.
The extension to $N>1$ is obvious. As a by product we observe that the vacuum wave function transforms as a scalar field
\be
O_0(\bm{k}) &\to& O_0(\bm{\Lambda^{-1} k}),\\
O_1(\bm{R}) &\to& O_1(\bm{\Lambda^{-1} R}),
\ee
implying
\be
Z_0(\bm{k}) &\to& Z_0(\bm{\Lambda^{-1} k}),\\
Z_1(\bm{R}) &\to& Z_1(\bm{\Lambda^{-1} R}),
\ee
and thus $Z=\max_{\bm k}\{Z_0(\bm{k})\}$ is SL(2,C) invariant, as stated in the preceding sections.

\subsection{Invariance of $H_1$}

The interaction Hamiltonian
\be
H_1(\tau) &=& \int d\tilde x_\tau J^a(x_\tau)A_a(x_\tau,N)= H_1(\tau,J,N)
\ee
is invariant
\be
U(\Lambda,0,N)^{\dag}H_1(\tau,J',N)U(\Lambda,0,N)=H_1(\tau,J,N).
\ee
Analogously, if $U_1(\tau,\tau_0,J,N)$ is the evolution operator whose generator is given by $H_1(\tau,J,N)$, then
\be
U(\Lambda,0,N)^{\dag}U_1(\tau,\tau_0,J',N)U(\Lambda,0,N)=U_1(\tau,\tau_0,J,N).
\ee
\subsection{Covariance of photon statistics}

Formulas (\ref{U-R}) and (\ref{L=J}) imply that the projector (\ref{Proj}) commutes with $U(\Lambda,y,N)$. Therefore,
\be
U_1(\tau,\tau_0,J',N)^{\dag}\Pi(n_+,n_-)U_1(\tau,\tau_0,J',N)
&=&
U(\Lambda^{-1},0,N)^{\dag}U_1(\tau,\tau_0,J,N)^{\dag}\Pi(n_+,n_-)U_1(\tau,\tau_0,J,N)U(\Lambda^{-1},0,N)\nonumber\\
\ee
The latter means that currents $J^a(x)$, $J'^a(x)$, related by a Lorentz transformation, produce different photon statistics, but the differences reduce to different forms of Lorentz-transformed vacuum probabilities: $Z_0(\bm k)$, $Z_1(\bm R)$, and $Z_0(\bm{\Lambda  k})$, $Z_1(\bm{\Lambda R})$.
In particular, if $J^a(x)$ corresponds to a pointlike charge moving with four-velocity $u_a$,
and $J'^a(x)$ describes an analogous charge moving with $u'_a=\Lambda_a{^b}u_b$, then vacuum-to-vacuum probabilities read, respectively,
(\ref{C -1}) and
\be
C'(-1,\tau,\infty)
&=&
\exp \Bigg(-q_{\rm ren}^2 \int d\tilde k{\,}
\chi(\bm{ \Lambda k})
\frac{\sin^2 (k\cdot u\tau/2)}{(k\cdot u/2)^2}
u^au^b
\Big(
\langle x_a(\bm k)x_b(\bm k)\rangle
+
\langle y_a(\bm k)y_b(\bm k)\rangle
\Big)
\Bigg).\label{C' -1}
\ee

\section{Closing remarks: Physical and formal structure of quantization in $N$-representations}

Quantum theory of light parametrized by $N$ seems to violate many apparently {\it sine qua non\/} standards imposed on present-day field theory, and yet the end result is the one one should expect: As $N$ increases the predictions converge to regularized versions of the standard divergent formulas. The fact that the weak limit $N\to\infty$ plays a role of a correspondence principle with the standard regularized theory is not accidental and was analyzed in the earlier papers \cite{I,II,III,CN,V}. So let us discuss in more detail also other important similarities and differences with respect to the standard formalism.

\subsection{Generator of evolution versus classical Noether invariant}

The Hamiltonian (\ref{H1 tau}) is a quantized version of the interaction part of a classical Noether invariant derived from the standard minimal-coupling Lagrangian density ${\cal L}(x)$. In classical theory one starts with the action $\int_{\cal O}d^4x {\cal L}(x)$, where the integration region is contained between two hyperboloids $\Sigma_{\tau_0}$ and $\Sigma_{\tau}$. The action is invariant under $(\sqrt{\tau^2+\bm x^2},\bm x)\to (\sqrt{(\tau+\epsilon)^2+\bm x^2},\bm x)$. Noether's theorem then leads to the Hamiltonian whose interaction part coincides with (\ref{H1 tau}). It is important that our (\ref{H1 tau}) is constructed from free fields evaluated on $\Sigma_{\tau}$. Let us stress that this is not exactly the standard interaction-picture since the measure $d\tilde x_\tau= d^3x/\sqrt{1+\bm x^2/\tau^2}$ depends on $\tau$. This is why we state at the end of Section III that this is how we {\it define\/} the dynamical system. The denominator in $d\tilde x_\tau$ shows that differences between $d\tilde x_\tau$ and $d^3x$ are negligible in present-day Earth-scale experiments (recall that $\tau$ is of the order of the age of the Universe). The fact that (\ref{H1 tau}) leads to the correct form of electrodynamics suggests an extension of our field quantization paradigm to other gauge theories in Minkowski space background: Take free fields in Minkowski space, define potentials in terms of free fields, and couple with an appropriate current, also evaluated in terms of free fields. This should be applicable, in particular, to the standard model and gauge theories of gravitation.

\subsection{Four-potential, gauge freedom, and indefinite-metric theories}

The difference between our formalism and those with indefinite metric lies in the fact that the positive-frequency part of free fields in momentum space contains both creation and annihilation operators. Their number depends on the signature of spacetime. Since the Minkowski space is $1+3$ dimensional the corresponding part of the potential contains one creation operator and three annihilation operators. In this way we obtain the potential which is Hermitian and thus the Hamiltonian generates a unitary dynamics. There are no negative norm states. Due to our choice of circular polarization vectors the Lorentz transformations are implemented in a unitary way, but the momentum space operators split into massless spin-1 and spin-0 representations. This does not contradict the fact that the position-space four-potential transforms as a four-vector field. Lorentz transformations do not change the form of the four-potential and thus do not change gauge. This form of ``gauge" is a consequence of the {\it geometric\/} condition (\ref{eq 1'}) --- the choice of explicit spin-frames satisfying (\ref{eq 1'}) is the only freedom left in our formalism. In this sense the four-potential is no longer an ordinary gauge field, in spite of the fact that the Hamiltonian was obtained by quantization of a gauge field.

The additional two scalar massless fields satisfy all the standards we impose on physical fields. We therefore predict that electrodynamics contains, in addition to the spin-1 massless photons, two massless scalar fields. The energy of the field with the index 3 is nonnegative, whereas the field indexed by 0 has non-positive free-field energy. The assumption that optical detectors react only to the spin-1 part turns the two scalar fields into a kind of dark matter. The electromagnetic vacuum is annihilated by annihilation operators indexed by 1 and 2, but in principle may contain the two types of scalar particles. In this sense what we call the electromagnetic vacuum may be a state of either positive or negative energy associated with the scalar particles. The choice of {\it exact\/} vacuum initial condition at $\tau=0$ seems natural, but photon statistics would be unchanged even if we started with a non-zero number of scalar particles.

The internal gauge group U(1) is implicitly present in the ambiguity of flag-plane $\pi_A(\bm k)\to e^{i\theta(\bm k)}\pi_A(\bm k)$.
As shown in \cite{H} this type of flag ambiguity when applied to massive fields implies SU$(n)$ gauge groups. So the flag ambiguity is indeed the gauge freedom in the sense used in particle physics.

\subsection{Link of COM representations to twistors}

Let us consider a twistor $(\pi_A,\omega^A)$ where $\omega^A=\omega^A(x,\pi)=\omega^A(0)+x^{AA'}\bar\pi_{A'}$. For massless particles one finds that
$\pi_A\bar\pi_{A'}=p^a$ is their momentum \cite{PR2}, and we know \cite{PR} that the inverse relation leads to $\pi_A=\pi_A(p)$, up to a flag-plane.
A twistor thus depends on momentum $p^a$ and position $x^a$ in the characteristic way: $\pi_A=\pi_A(p)$, $\omega_A=\omega_A(x,p)$.
The new spinor field $\Lambda{_A}{^B}\pi_B(\Lambda^{-1}p)$ has the same flag-pole as $\pi_A(p)$ an thus differs at most by the flag-plane.  Let
$\Lambda{_A}{^B}\pi_B(\Lambda^{-1}p)=e^{i\theta}\pi_A(p)$. The inhomogeneous part of the twistor transforms by
\be
x^{AA'}\bar\pi_{A'}(p)\to (\Lambda^{-1}x)^{AA'}\overline{(\Lambda\pi)}{_{A'}}(\Lambda^{-1}p)=e^{-i\theta}x^{AA'}\bar\pi_{A'}(p)
\ee
which is equivalent to (\ref{eq 1'}). It follows that the twistor equation is a particular form of (\ref{eq 1'}). Twistor is not the most general solution of (\ref{eq 1'}) since one can multiply twistors by any function of the argument $p^ax_a$ and yet obtain (\ref{eq 1'}). What we in fact have is a twistor equation constrained by $\omega_A\pi^A=1$. We have not studied the constrained equation in its full generality.

\subsection{Weak cutoff, weak law of large numbers, correspondence principle}

The cutoff function does not appear at the level of operators, but enters as a consequence of evaluating averages in states. The representations of CCR, the Hamiltonian, and the evolution operator are all independent of any cutoff. Instead, in the places where cutoff is typically inserted by hand we encounter the central element $I(\bm \kappa,N)$, which may be regarded as kind of cutoff operator. This is why, following David Finkelstein \cite{Finkelstein}, we speak of ``regularization by quantization". The link between ``cutoff operators" and the true cutoff function is: $\langle 0,N|I(\bm k,N)|0,N\rangle=\langle 0,1|I(\bm k,1)|0,1\rangle=Z\chi(\bm k)$,
$\int d\tilde R\langle 0,N|I(\bm R,\bm k,N)|0,N\rangle=\int d\tilde R\langle 0,1|I(\bm R,\bm k,1)|0,1\rangle=Z\chi(\bm k)$. In irreducible representations we get $\langle 0|I(\bm k)|0\rangle=Z$, a fact that explains how to compare predictions of theories based on $N$-representations with those of the irreducible representations of CCR: Take the limit $N\to\infty$ and replace $\chi(\bm k)$ by 1. For $N=1$ the central element $I(\bm \kappa,1)$ is a projector. For $N>1$ we have
\be
I(\bm \kappa,N)
&=&
\frac{1}{N}\Big(I(\bm \kappa,1)\otimes I \otimes\dots \otimes I+\dots+I \otimes \dots \otimes I\otimes  I(\bm \kappa,1)\Big)
\ee
which means that $I(\bm \kappa,N)$ is the so-called frequency-of-successes operator know from quantum laws of large numbers \cite{LLN1,LLN2,LLN3,LLN4} and whose eigenvalues are $s/N$, $s=0,1,\dots,N$. This explains why in the weak limit $N\to\infty$ we effectively obtain $Z(\bm \kappa)$ in all those places where $I(\bm \kappa,N)$ occurred. It should be stressed that the cutoff $\chi(\bm k)$ is an inherent part of the prediction and reflects the physical structure of the Bose-Einstein condensate $|0,N\rangle$, playing the role of vacuum. There is no reason to remove $\chi(\bm k)$ from the formulas. This should be contrasted with the various approaches to the infrared problem one finds in the literature, where the cutoff has no real physical justification and has to be removed at the end of the calculation \cite{JR,YFS,Ch,K,FK,F}.

The difference between our type of cutoff and the strong one typically employed in quantum optical models is similar to that between {\it trace\/} of an operator, and its matrix element. Indeed, let us consider the Hamiltonian $H=\sum_{n=1}^{\infty}n|n\rangle \langle n|$. Its trace is divergent,
$\Tr H=\sum_{n=1}^{\infty}n$, but its matrix element $\langle \psi|H|\psi\rangle=\sum_{n=1}^{\infty}n|\psi_n|^2$ is finite in an appropriate domain.
But let us note that that average is indistinguishable from the trace of the regularized operator $H=\sum_{n=1}^{\infty}|\psi_n|^2n|n\rangle \langle n|$. The latter is an example of ``strong" regularization we know from standard approaches to field quantization, the former is what happens in $N$-representations as a consequence of the weak law of large numbers. This is why our ``regularization" cannot be seen in spectra of Hamiltonians.

\subsection{Non-locality versus causality}

The solution of Heisenberg's equation of motion satisfies also Maxwell's equations, but with an {\it operator\/} current that is not identical to the classical current occurring in the Hamiltonian. The two currents are different in $N$-representations because of the nontrivial structure of the central elements.
A coherent state average of a solution can be regarded as a classical limit of the theory. This classical solution satisfies the classical Maxwell equations but, of course, with appropriately modified current.

Quantum fields quantized in $N$-representations are not local, in the meaning of this term employed in axiomatic quantum field theory \cite{Haag}, but it does not mean they are acausal: Solutions of the Heisenberg equation satisfy causal Maxwell equations. What we call nonlocality means here effectively that pointlike charges behave as if they were extended. Our quantum field theory peacefully coexists with causality, but differs from the other approaches to nonlocal fields one finds in the literature \cite{Efimov,NL2,NL3,NL4,NL5,NL6}. The approach based on $N$-representations should not be confused with generalized free fields \cite{BLT} where the right-hand-sides of CCR involve Casimir invariants of relativistic symmetries.

\section{Appendices}

\subsection{Tetrads and spin-frames --- Notation and basic technicalities}
Our notation is essentially an appropriately adapted version of the abstract-index convention of Newman, Penrose, and Rindler \cite{PR}. The boldface indices
$\textbf{a}$, $\textbf{A}$, take numerical values $0$, $1$, $2$, $3$, and $0$, $1$, respectively, and are related to a
concrete choice of basis. The italics $a$, $A$ are abstract indices and specify types of objects.
We work in Minkowski space of signature $(+,-,-,-)$. The metric tensor is denoted by
$g_{ab}$. $g_{\textbf{ab}}$ and $g^{\textbf{ab}}$ are the matrices diag$(+,-,-,-)$.
Minkowski tetrads, indexed by indices that are
partly boldfaced and partly italic, say
$g_{a}^{~\textbf{a}}$, consist of four four-vectors (or four-vector fields)
$g_{a}^{~0}$, $g_{a}^{~1}$, $g_{a}^{~2}$, $g_{a}^{~3}$. We will work with three types of tetrads, $g_{a}{^{\bf a}}$, $g_{a}{^{\bf a}}(\bm k)$, and $g_{a}{^{\bf a}}(\bm R,\bm k)$. The momentum independent tetrad
$g_{a}^{~\bf a}$ satisfies $k^0=|\bm k|=k^a g_{a}^{~0}$,
$k^1=k^a g_{a}^{~1}$, $k^2=k^a g_{a}^{~2}$, $k^3=k^a g_{a}^{~3}$,
and defines decomposition of four-momentum into energy and three-momentum employed, for example, in the invariant measure
$d\tilde k=(2\pi)^{-3}d^3k/(2 |\bm k|)$.
The four momentum $k^{a}=k^{a}(\bm k)$ can be written in spinor notation as
$k^{a}(\bm k)=\pi^{A}(\bm k)\bar\pi^{A'}(\bm k)$,
where $\pi^{A}(\bm k)$  is a spinor field defined by $k^{a}(\bm k)$
up to a phase factor. For any $\pi^{A}(\bm k)$ there exists
another spinor field $\omega^{A}(\bm k)$
satisfying the spin-frame condition
$\omega_{A}(\bm k)\pi^{A}(\bm k)=1$. The spin-frame condition does not uniquely determine $\omega_{A}(\bm k)$ on the basis of $\pi_{A}(\bm k)$, and the associated freedom is of a gauge type.
Indices $a$ and $\bf a$ can be raised and lowered by means of the
metric tensors $g_{ab}$, $g^{ab}$, $g_{\bf ab}$, $g^{\bf ab}$. The link between metric tensors and Minkowski tetrads is
\be
g_{ab} &=& g_{a}^{~\bf a}(\bm k)g_{b}^{~\bf b}(\bm k)g_{\bf ab}=g_{a}^{~\bf a}(\bm R,\bm k)g_{b}^{~\bf b}(\bm R,\bm k)g_{\bf ab}
=g_{a}^{~\bf a}g_{b}^{~\bf b}g_{\bf ab},\\
g^{\bf ab} &=& g_{a}^{~\bf a}(\bm k)g_{b}^{~\bf b}(\bm k)g^{ab}=g_{a}^{~\bf a}(\bm R,\bm k)g_{b}^{~\bf b}(\bm R,\bm k)g^{ab}=
g_{a}^{~\bf a}g_{b}^{~\bf b}g^{ab}.
\ee
Null tetrads will be indexed by indices that are partly boldface-primed and partly italic:
$g^{a}_{~~\textbf{b}'}  $
It is important to distinguish between $\bf a$ and $\bf a'$, and we will employ the convention where
${\bf a'}=00',01',10',11'$,
with $g_{\bf a'b'}=\ve_{\bf AB}\ve_{\bf A'B'}$, $g^{\bf a'b'}=\ve^{\bf AB}\ve^{\bf A'B'}$.  We have to raise
and lower indices  $\bf AA'$ by means of
$\ve_{\bf AB}\ve_{\bf A'B'}$. The null tetrad associated with spin-frames can be written as
\begin{eqnarray}
g^{a}{_{\textbf{b}'}}=
\left(
\begin{array}{c}
g^{a}_{~~00'}\\
g^{a}_{~~01'}\\
g^{a}_{~~10'}\\
g^{a}_{~~11'}
\end{array}
\right) =
\left(
\begin{array}{c}
\varepsilon^{A}_{~~0}\varepsilon^{A'}_{~~0'}\\
\varepsilon^{A}_{~~0}\varepsilon^{A'}_{~~1'}\\
\varepsilon^{A}_{~~1}\varepsilon^{A'}_{~~0'}\\
\varepsilon^{A}_{~~1}\varepsilon^{A'}_{~~1'}
\end{array}
\right)=
\left(
\begin{array}{c}
\omega^{A}\bar\omega^{A'}\\
\omega^{A}\bar\pi^{A'}\\
\pi^{A}\bar\omega^{A'}\\
\pi^{A}\bar\pi^{A'}
\end{array}
\right)
=
\left(
\begin{array}{c}
\omega^{a}\\
m^{a}\\
\bar{m}^{a}\\
k^{a}
\end{array}
\right)\label{4}
\end{eqnarray}
and dually
\begin{eqnarray}
g_{a}{^{\textbf{b}'}}=
 \left(
\begin{array}{c}
g_{a}^{~~00'}\\
g_{a}^{~~01'}\\
g_{a}^{~~10'}\\
g_{a}^{~~11'}
\end{array}
\right) =
\left(
\begin{array}{c}
\varepsilon_{A}^{~~0}\varepsilon_{A'}^{~~0'}\\
\varepsilon_{A}^{~~0}\varepsilon_{A'}^{~~1'}\\
\varepsilon_{A}^{~~1}\varepsilon_{A'}^{~~0'}\\
\varepsilon_{A}^{~~1}\varepsilon_{A'}^{~~1'}
\end{array}
\right)
=
\left(
\begin{array}{c}
\pi_{A}\bar\pi_{A'}\\
-\pi_{A}\bar\omega_{A'}\\
-\omega_{A}\bar\pi_{A'}\\
\omega_{A}\bar\omega_{A'}
\end{array}
\right)
=
\left(
\begin{array}{c}
k_{a}\\
-\bar{m}_{a}\\
-m_{a}\\
\omega_{a}
\end{array}
\right)\label{5}
\end{eqnarray}
There is a relation between the Minkowski tetrad, indexed by indices that are
partly boldface and partly italic, and  the null tetrad (we skip the arguments $\bm k$ and $\bm R$, but the formulas are valid for all the tetrads of interest)
\begin{eqnarray}
g^{a}{_{\textbf{a}}}
=
g_{\textbf{a}\textbf{b}'}g^{a\textbf{b}'}
=
g_{\textbf{a}}{^{\textbf{b}'}}g^{a}{_{\textbf{b}'}}
=
g_{\textbf{a}}{^{\textbf{BB}'}}g^{a}{_{\textbf{BB}'}}
\end{eqnarray}
\begin{eqnarray}
\left(
\begin{array}{c}
t^{a}\\
x^{a}\\
y^{a}\\
z^{a}
\end{array}
\right) = \left(
\begin{array}{c}
g^{a}_{~~0}\\
g^{a}_{~~1}\\
g^{a}_{~~2}\\
g^{a}_{~~3}
\end{array}
\right) = \frac{1}{\sqrt{2}} \left(
\begin{array}{cccc}
1&0&0&1\\
0&1&1&0\\
0&i&-i&0\\
1&0&0&-1
\end{array}
\right) \left(
\begin{array}{c}
\omega^{A}\bar\omega^{A'}\\
\omega^{A}\bar\pi^{A'}\\
\pi^{A}\bar\omega^{A'}\\
\pi^{A}\bar\pi^{A'}
\end{array}
\right)
=
 \frac{1}{\sqrt{2}}
 \left(
\begin{array}{c}
\omega^{a}+k^{a}\\
m^{a}+\bar{m}^{a}\\
i m^{a} - i\bar{m}^{a}\\
\omega^{a}-k^{a}
\end{array}
\right)
\end{eqnarray}
and dually
\begin{eqnarray}
g_{a}{^{\textbf{a}}}
=
g^{\textbf{a}\textbf{b}'}g_{a\textbf{b}'}
=
g^{\textbf{a}}{_{\textbf{b}'}}g_{a}{^{\textbf{b}'}}
=
g^{\textbf{a}}{_{\textbf{BB}'}}g_{a}{^{\textbf{BB}'}}
\end{eqnarray}
\begin{eqnarray}
\left(
\begin{array}{c}
t_{a}\\
-x_{a}\\
-y_{a}\\
-z_{a}
\end{array}
\right) = \left(
\begin{array}{c}
g_{a}^{~~0}\\
g_{a}^{~~1}\\
g_{a}^{~~2}\\
g_{a}^{~~3}
\end{array}
\right) = \frac{1}{\sqrt{2}} \left(
\begin{array}{cccc}
1&0&0&1\\
0&1&1&0\\
0&-i&i&0\\
1&0&0&-1
\end{array}
\right) \left(
\begin{array}{c}
\pi_{A}\bar\pi_{A'}\\
-\pi_{A}\bar\omega_{A'}\\
-\omega_{A}\bar\pi_{A'}\\
\omega_{A}\bar\omega_{A'}
\end{array}
\right)=
 \frac{1}{\sqrt{2}}
 \left(
\begin{array}{c}
k_{a}+\omega_{a}\\
-\bar{m}_{a}-m_{a}\\
i \bar{m}_{a} -i m_{a}\\
k_{a}-\omega_{a}
\end{array}
\right)
\end{eqnarray}
Here the $g$s with partly boldface  and partly boldface-primed
indices are the Infeld-van der Waerden symbols
\be
g^{\textbf{a}}{_{\textbf{b}'}} &=&
\left(
\begin{array}{cccc}
g^{0}{_{00'}} & g^{0}{_{01'}} & g^{0}{_{10'}} & g^{0}{_{11'}} \\
g^{1}{_{00'}} & g^{1}{_{01'}} & g^{1}{_{10'}} & g^{1}{_{11'}} \\
g^{2}{_{00'}} & g^{2}{_{01'}} & g^{2}{_{10'}} & g^{2}{_{11'}} \\
g^{3}{_{00'}} & g^{3}{_{01'}} & g^{3}{_{10'}} & g^{3}{_{11'}}
\end{array}
\right)
=
\frac{1}{\sqrt{2}}
\left(
\begin{array}{cccc}
1&0&0&1\\
0&1&1&0\\
0&-i&i&0\\
1&0&0&-1\\
\end{array}
\right),\\
g_{\textbf{a}}{^{\textbf{b}'}}
&=&
\left(
\begin{array}{cccc}
g_{0}{^{00'}} & g_{0}{^{01'}} & g_{0}{^{10'}} & g_{0}{^{11'}} \\
g_{1}{^{00'}} & g_{1}{^{01'}} & g_{1}{^{10'}} & g_{1}{^{11'}} \\
g_{2}{^{00'}} & g_{2}{^{01'}} & g_{2}{^{10'}} & g_{2}{^{11'}} \\
g_{3}{^{00'}} & g_{3}{^{01'}} & g_{3}{^{10'}} & g_{3}{^{11'}}
\end{array}
\right)
=
\frac{1}{\sqrt{2}} \left(
\begin{array}{cccc}
1&0&0&1\\
0&1&1&0\\
0&i&-i&0\\
1&0&0&-1
\end{array}
\right).
\ee
Let us introduce the SO(1,3) matrices
\be
L_{\textbf{a}}{^\textbf{b}}(\Lambda, \bm k)
&=&
g_{~\textbf{a}}^{a}(\bm k)\Lambda{_a}{^b}g^{~\textbf{b}}_{b}(\boldsymbol{\Lambda^{-1}k})
=
g_{~\textbf{a}}^{a}(\bm k)\Lambda g^{~\textbf{b}}_{a}(\boldsymbol{k})
\label{LL},\\
L_{\textbf{a}}{^\textbf{b}}(\Lambda, \bm R,\bm k)
&=&
g_{~\textbf{a}}^{a}(\bm R,\bm k)\Lambda{_a}{^b}g^{~\textbf{b}}_{b}(\bm{\Lambda^{-1}R},\boldsymbol{\Lambda^{-1}k})
=
g_{~\textbf{a}}^{a}(\bm R,\bm k)\Lambda g^{~\textbf{b}}_{a}(\bm R,\boldsymbol{k})
\label{LL-R},
\ee
where $\boldsymbol{\Lambda^{-1} k}$, $\boldsymbol{\Lambda^{-1} R}$ are the spacelike parts of the four vectors $\Lambda^{-1}{_a}{^b}k_{b}(\bm k)$,
$\Lambda^{-1}{_a}{^b}R_{b}$, respectively, and $\Lambda g^{~\textbf{b}}_{a}(\boldsymbol{k})=\Lambda{_a}{^b}g^{~\textbf{b}}_{b}(\boldsymbol{\Lambda^{-1}k})$,
$\Lambda g^{~\textbf{b}}_{a}(\bm R,\boldsymbol{k})=
\Lambda{_a}{^b}g^{~\textbf{b}}_{b}(\bm{\Lambda^{-1}R},\boldsymbol{\Lambda^{-1}k})$.

In order to derive an explicit form of (\ref{LL}) we have to control transformation properties of the tetrad field.
The basic assumption is that the spin-frames are spinor fields, i.e. transform by
\be
\pi_{A}(\bm k)&\mapsto& \Lambda\pi_{A}(\bm k)=
\Lambda_{A}^{~~B}\pi_{B}(\boldsymbol{\Lambda^{-1}k}),\\
\omega_{A}(\bm k) &\mapsto &
\Lambda\omega_{A}(\bm k)=
\Lambda_{A}^{~~B}\omega_{B}(\boldsymbol{\Lambda^{-1}k}),\\
\pi_{A}(\bm R,\bm k)&\mapsto& \Lambda\pi_{A}(\bm k)=
\Lambda_{A}^{~~B}\pi_{B}(\bm{\Lambda^{-1}R},\boldsymbol{\Lambda^{-1}k}),\\
\omega_{A}(\bm R,\bm k) &\mapsto &
\Lambda\omega_{A}(\bm k)=
\Lambda_{A}^{~~B}\omega_{B}(\bm{\Lambda^{-1}R},\boldsymbol{\Lambda^{-1}k}).
\ee
Here $\Lambda_{A}{^{B}}$ denotes
an unprimed SL(2,C) transformation corresponding to $\Lambda_{a}{^{b}}\in$ SO(1,3).
Since
$\Lambda\pi_{A}(\bm k)\overline{\Lambda\pi}{_{A'}}(\bm k)=\pi_{A}(\bm k)\bar\pi_{A'}(\bm k)=\pi_{A}(\bm R,\bm k)\bar\pi_{A'}(\bm R,\bm k)
=\Lambda\pi_{A}(\bm R,\bm k)\overline{\Lambda\pi}{_{A'}}(\bm R,\bm k)$
is satisfied by definition of $\pi_{A}(\bm k)$ and $\pi_{A}(\bm R,\bm k)$
one finds
\be
\Lambda\pi_{A}(\bm k) &=&
e^{-i\Theta(\Lambda,\bm k)}\pi_{A}(\bm k),\\
\Lambda\pi_{A}(\bm R,\bm k) &=&
e^{-i\Theta(\Lambda,\bm k)}\pi_{A}(\bm R,\bm k).
\ee
The angle $\Theta(\Lambda,\bm k)$ is the spin-1/2, zero-mass Wigner phase known from unitary representations of the Poincar\'e group (purely spinorial proof of this statement can be found in \cite{MC99}), and this is why it does not depend on $\bm R$.
Conservation of the spin-frame condition,
$
\Lambda\omega_{A}(\bm k)\Lambda\pi^{A}(\bm k)= \omega_{A}(\bm k)\pi^{A}(\bm k)=1,
$
implies
\begin{equation}
\Lambda\omega_{A}(\bm k)=
e^{i\Theta(\Lambda,\bm k)}\big(\omega_{A}(\bm k)+\phi(\Lambda,\bm k)\pi_{A}(\bm k)\big).
\end{equation}
$\phi(\Lambda,\bm k) = |\phi(\Lambda,\bm k)| e^{i\xi(\Lambda,\bm k)}$
is a complex number that depends on the explicit form of $\omega_{A}(\bm k)$ (which is non-unique) and thus is a quantity of a gauge type. Obviously,
\be
e^{i\Theta(\Lambda,\bm k)}
&=&
\pi^{A}(\bm k)
\Lambda\omega_{A}(\bm k),\\
\phi(\Lambda,\bm k)
&=&
e^{-i\Theta(\Lambda,\bm k)}\omega_{A}(\bm k)
\Lambda\omega^{A}(\bm k).
\ee
In terms of this parametrization we find
\be
\left(
\begin{array}{c}
\Lambda t^{a}(\bm k)\\
\Lambda x^{a}(\bm k)\\
\Lambda y^{a}(\bm k)\\
\Lambda z^{a}(\bm k)
\end{array}
\right)
&=&
\frac{1}{\sqrt{2}} \left(
\begin{array}{cccc}
1&0&0&1\\
0&1&1&0\\
0&i&-i&0\\
1&0&0&-1
\end{array}
\right) \left(
\begin{array}{c}
\Lambda \omega^{A}(\bm k)\overline{\Lambda \omega}{^{A'}}(\bm k)\\
\Lambda \omega^{A}(\bm k)\overline{\Lambda \pi}{^{A'}}(\bm k)\\
\Lambda \pi^{A}(\bm k)\overline{\Lambda \omega}{^{A'}}(\bm k)\\
\Lambda \pi^{A}(\bm k)\overline{\Lambda \pi}{^{A'}}(\bm k)
\end{array}
\right)\label{LLL}\\
&=&
\left(
\begin{array}{c}
(1+|\phi|^2/2)t^a
+|\phi|\cos\xi x^a
-|\phi|\sin\xi y^a
-|\phi|^2/2 z^a
\\
|\phi|\cos (2\Theta+\xi) t^a
+\cos 2\Theta x^a
+\sin 2\Theta y^a
-|\phi|\cos (2\Theta+\xi)z^a
\\
-|\phi|\sin (2\Theta+\xi) t^a
-\sin 2\Theta x^a
+\cos 2\Theta y^a
+|\phi|\sin (2\Theta+\xi)z^a
\\
|\phi|^2/2t^a
+
|\phi|\cos \xi x^a
-
|\phi|\sin \xi y^a
+
(1-|\phi|^2/2)z^a
\end{array}
\right)\label{LLLL}
\ee
and
\be
L_{\textbf{a}}{^\textbf{b}}(\Lambda, \bm k)
&=&
\left(
\begin{array}{cccc}
t_a(\bm k)\Lambda t^a(\bm k) &
-t_a(\bm k)\Lambda x^a(\bm k) &
-t_a(\bm k)\Lambda y^a(\bm k) &
-t_a(\bm k)\Lambda z^a(\bm k) \\
x_a(\bm k)\Lambda t^a(\bm k) &
-x_a(\bm k)\Lambda x^a(\bm k) &
-x_a(\bm k)\Lambda y^a(\bm k) &
-x_a(\bm k)\Lambda z^a(\bm k) \\
y_a(\bm k)\Lambda t^a(\bm k) &
-y_a(\bm k)\Lambda x^a(\bm k) &
-y_a(\bm k)\Lambda y^a(\bm k) &
-y_a(\bm k)\Lambda z^a(\bm k) \\
z_a(\bm k)\Lambda t^a(\bm k) &
-z_a(\bm k)\Lambda x^a(\bm k) &
-z_a(\bm k)\Lambda y^a(\bm k) &
-z_a(\bm k)\Lambda z^a(\bm k)
\end{array}
\right)\\
&=&
\left(
\begin{array}{cccc}
1+\frac{|\phi|^2}{2} & -|\phi|\cos(\xi+2\Theta) &
|\phi|\sin(\xi+2\Theta) &
-\frac{|\phi|^2}{2}\\
-|\phi|\cos \xi
& \cos 2\Theta & -\sin 2\Theta &
|\phi|\cos \xi\\
|\phi|\sin \xi &
\sin 2\Theta & \cos 2\Theta &
-|\phi|\sin \xi\\
\frac{|\phi|^{2}}{2}
& -|\phi|\cos(\xi+2\Theta)
& |\phi|\sin
(\xi+2\Theta) & 1-\frac{|\phi|^{2}}{2}
\end{array}
\right).\label{LLLLL}
\ee
Here, of course,  $\Theta=\Theta(\Lambda,\bm k)$, $|\phi|=|\phi(\Lambda,\bm k)|$, and $\xi=\xi(\Lambda,\bm k)$. Just for completness let us note that $L_{\textbf{a}}{^\textbf{b}}(\Lambda, \bm k)$ corresponds to the SL(2,C) matrix
\be
L_{\textbf{A}}{^{\textbf{B}}}(\Lambda, \bm k)
&=&
\varepsilon_{\textbf{A}}^{~~A}(\bm k)
\Lambda_{A}^{~~B}
\varepsilon^{~~\textbf{B}}_{B}(\boldsymbol{\Lambda^{-1}k})
=
\varepsilon_{\textbf{A}}^{~~A}(\bm k)
\Lambda\varepsilon^{~~\textbf{B}}_{A}(\bm k)\nonumber\\
&=&
\left(
\begin{array}{cc}
\omega_{A}(\bm k)
\Lambda
\pi^{A}(\bm k)&
\omega^{A}(\bm k)
\Lambda
\omega_{A}(\bm k)\\
0&
\pi^{A}(\bm k)
\Lambda
\omega_{A}(\bm k)
\end{array}
\right)
= \left(
\begin{array}{cc}
e^{-i\Theta(\Lambda,\bm k)}
& -\phi(\Lambda,\bm k)e^{i\Theta(\Lambda,\bm k)}\\
0 & e^{i\Theta(\Lambda,\bm k)}\\
\end{array}
\right)
\label{17}.
\end{eqnarray}
Now let us consider the spin-frame (\ref{pppp})--(\ref{oooo}). We find
\be
\left(
\begin{array}{c}
\Lambda t^{a}(\bm R,\bm k)\\
\Lambda x^{a}(\bm R,\bm k)\\
\Lambda y^{a}(\bm R,\bm k)\\
\Lambda z^{a}(\bm R,\bm k)
\end{array}
\right)
&=&
\frac{1}{\sqrt{2}} \left(
\begin{array}{cccc}
1&0&0&1\\
0&1&1&0\\
0&i&-i&0\\
1&0&0&-1
\end{array}
\right) \left(
\begin{array}{c}
\Lambda \omega^{A}(\bm R,\bm k)\overline{\Lambda \omega}{^{A'}}(\bm R,\bm k)\\
\Lambda \omega^{A}(\bm R,\bm k)\overline{\Lambda \pi}{^{A'}}(\bm R,\bm k)\\
\Lambda \pi^{A}(\bm R,\bm k)\overline{\Lambda \omega}{^{A'}}(\bm R,\bm k)\\
\Lambda \pi^{A}(\bm R,\bm k)\overline{\Lambda \pi}{^{A'}}(\bm R,\bm k)
\end{array}
\right)\label{LLL-R}\\
&=&
\left(
\begin{array}{c}
t^a(\bm R,\bm k)
\\
\cos 2\Theta(\Lambda,\bm k) x^a(\bm R,\bm k)
+\sin 2\Theta(\Lambda,\bm k) y^a(\bm R,\bm k)
\\
-\sin 2\Theta(\Lambda,\bm k) x^a(\bm R,\bm k)
+\cos 2\Theta(\Lambda,\bm k) y^a(\bm R,\bm k)
\\
z^a(\bm R,\bm k)
\end{array}
\right)\label{LLLL-R}
\ee
and
\be
L_{\textbf{a}}{^\textbf{b}}(\Lambda,\bm R, \bm k)
&=&
\left(
\begin{array}{cccc}
t_a(\bm R,\bm k)\Lambda t^a(\bm R,\bm k) &
-t_a(\bm R,\bm k)\Lambda x^a(\bm R,\bm k) &
-t_a(\bm R,\bm k)\Lambda y^a(\bm R,\bm k) &
-t_a(\bm R,\bm k)\Lambda z^a(\bm R,\bm k) \\
x_a(\bm R,\bm k)\Lambda t^a(\bm R,\bm k) &
-x_a(\bm R,\bm k)\Lambda x^a(\bm R,\bm k) &
-x_a(\bm R,\bm k)\Lambda y^a(\bm R,\bm k) &
-x_a(\bm R,\bm k)\Lambda z^a(\bm R,\bm k) \\
y_a(\bm R,\bm k)\Lambda t^a(\bm R,\bm k) &
-y_a(\bm R,\bm k)\Lambda x^a(\bm R,\bm k) &
-y_a(\bm R,\bm k)\Lambda y^a(\bm R,\bm k) &
-y_a(\bm R,\bm k)\Lambda z^a(\bm R,\bm k) \\
z_a(\bm R,\bm k)\Lambda t^a(\bm R,\bm k) &
-z_a(\bm R,\bm k)\Lambda x^a(\bm R,\bm k) &
-z_a(\bm R,\bm k)\Lambda y^a(\bm R,\bm k) &
-z_a(\bm R,\bm k)\Lambda z^a(\bm R,\bm k)
\end{array}
\right)\\
&=&
\left(
\begin{array}{cccc}
1 & 0 &  0& 0\\
0 & \cos 2\Theta(\Lambda,\bm k) & -\sin 2\Theta(\Lambda,\bm k) & 0\\
0 & \sin 2\Theta(\Lambda,\bm k) & \cos 2\Theta(\Lambda,\bm k) & 0\\
0 & 0 & 0 & 1
\end{array}
\right).\label{LLLLL-R}
\ee

\subsection{Construction of $U(\Lambda,0,N)$}

Consider the following sequence of transformations (repeated indices $j$, $j'$ are summed from 1 to 3):
\be
\Lambda{_a}{^b}A_{b}(\Lambda^{-1}x,1)
&=&
i\int d\tilde k{\,}\left(
\Lambda{_a}{^b}g_{b}^{~j}(\bm k)a_j(\bm k,1)
+
\Lambda{_a}{^b}g_{b}^{~0}(\bm k)a_0(\bm k,1)^{\dagger}
\right) e^{-ik\cdot \Lambda^{-1}x}+{\rm H.c.}\label{T1}\\
&=&
i\int d\tilde k{\,}|\bm k\rangle\langle\bm k|\otimes
\left(
\Lambda{_a}{^b}g_{b}^{~j}(\bm k) a_j
+
\Lambda{_a}{^b}g_{b}^{~0}(\bm k) a_0^{\dagger}
\right) e^{-i\Lambda k\cdot x}+{\rm H.c.}\label{T2}\\
&=&
i\int d\tilde k{\,}\Lambda{_a}{^b}g_{b}^{~1}(\bm{\Lambda^{-1} k})
|\bm {\Lambda^{-1} k}\rangle\langle\bm {\Lambda^{-1} k}|\otimes a_1 e^{-ik\cdot x}+\dots\label{T3}\\
&=&
W(\Lambda)\Big(i\int d\tilde k{\,}\Lambda{_a}{^b}g_{b}^{~1}(\bm{\Lambda^{-1} k})
|\bm {k}\rangle\langle\bm {k}|\otimes a_1
e^{-ik\cdot x}+\dots\Big)W(\Lambda)^{\dag}\label{T4}\\
&=&
W(\Lambda)\Big(i\int d\tilde k{\,}g_{a}{^{\bf b}}(\bm{k})
\underbrace{g^{b}{_{\bf b}}(\bm{k})\Lambda{_b}{^c}g_{c}^{~1}(\bm{\Lambda^{-1} k})}
_{L{_{\bf b}}{^1}(\Lambda, \bm k)}
|\bm {k}\rangle\langle\bm {k}|\otimes a_1
e^{-ik\cdot x}+\dots\Big)W(\Lambda)^{\dag}\label{T5}\\
&=&
W(\Lambda)i\int d\tilde k{\,}g_{a}{^{0}}(\bm{k})
|\bm {k}\rangle\langle\bm {k}|\otimes \Big(
L{_{0}}{^0}(\Lambda, \bm k)a_0^{\dag}
+
L{_{0}}{^j}(\Lambda, \bm k)a_j
\Big)e^{-ik\cdot x}W(\Lambda)^{\dag}\nonumber\\
&\pp=&+
W(\Lambda)i\int d\tilde k{\,}g_{a}{^{j}}(\bm{k})
|\bm {k}\rangle\langle\bm {k}|\otimes \Big(
L{_{j}}{^0}(\Lambda, \bm k)a_0^{\dag}
+
L{_{j}}{^{j'}}(\Lambda, \bm k)a_{j'}
\Big)e^{-ik\cdot x}W(\Lambda)^{\dag}+{\rm H.c.}\label{T8}\\
&=&
W(\Lambda)i\int d\tilde k{\,}g_{a}{^{0}}(\bm{k})
|\bm {k}\rangle\langle\bm {k}|\otimes
V(\Lambda,\bm k)a_0^{\dag}V(\Lambda,\bm k)^{\dag}
e^{-ik\cdot x}W(\Lambda)^{\dag}\nonumber\\
&\pp=&+
W(\Lambda)i\int d\tilde k{\,}g_{a}{^{j}}(\bm{k})
|\bm {k}\rangle\langle\bm {k}|\otimes V(\Lambda,\bm k)a_jV(\Lambda,\bm k)^{\dag}
e^{-ik\cdot x}W(\Lambda)^{\dag}+{\rm H.c.}\label{T9}\\
&=&
W(\Lambda)V(\Lambda) A_{a}(x,1)V(\Lambda)^{\dag}W(\Lambda)^{\dag}
\ee
(\ref{T1})$\to$(\ref{T2}) follows from
$k\cdot \Lambda^{-1}x=\Lambda k\cdot x$ and the form of $N=1$ representation of CCR.
(\ref{T2})$\to$(\ref{T3}) is the change of variables under integral, employing Lorentz invariance of $d\tilde k{\,}$. In (\ref{T3})$\to$(\ref{T4}) we introduce the unitary operator
$W(\Lambda)\big(|\bm {k}\rangle\langle\bm k|\otimes 1\big)W(\Lambda)^{\dag}
=|\bm {\Lambda^{-1}k}\rangle \langle\bm {\Lambda^{-1}k}|\otimes 1$.
(\ref{T4})$\to$(\ref{T5}) uses the Minkowski-tetrad condition
$g_{a}{^{\bf b}}(\bm{k})g^{b}{_{\bf b}}(\bm{k})=g{_a}{^b}$ and defines the matrix (\ref{LL})
which connects two Minkowski tetrads and thus belongs to  SO(1,3).
The crucial element of the construction is (\ref{T8})$\to$(\ref{T9}), where
$V(\Lambda,\bm k)$ is unitary.
To understand why $V(\Lambda,\bm k)$
has to exist consider an arbitrary matrix $L{_{\bf a}}{^{\bf b}}$ satisfying the SO(1,3) condition
$
L{_{\bf a}}{^{\bf c}}L{_{\bf b}}{^{\bf d}}g_{\bf cd}=g_{\bf ab}
$
and define
\be
b_{0}^{\dagger}
&=&
L_{0}^{~0}a_{0}^{\dagger}+L_{0}^{~1}a_{1}+L_{0}^{~2}a_{2}+L_{0}^{~3}a_{3},
\\
b_{1}
&=&
L_{1}^{~0}a_{0}^{\dagger}+L_{1}^{~1}a_{1}+L_{1}^{~2}a_{2}+L_{1}^{~3}a_{3},
\\
b_{2}
&=&
L_{2}^{~0}a_{0}^{\dagger}+L_{2}^{~1}a_{1}+L_{2}^{~2}a_{2}+L_{2}^{~3}a_{3},
\\
b_{3}
&=&
L_{3}^{~0}a_{0}^{\dagger}+L_{3}^{~1}a_{1}+L_{3}^{~2}a_{2}+L_{3}^{~3}a_{3}.
\ee
If $[a_{\bf a},a_{\bf b}^{\dag}]=\delta_{\bf ab}1$ then
$[b_{\bf a},b_{\bf b}^{\dag}]=\delta_{\bf ab}1$, as can be checked by explicit calculation employing $g_{\bf ab}={\rm diag}(+1,-1,-1,-1)$ (this trick can be generalized to any dimension and any signature). Now, by a well known theorem there exists a unitary
$V$  such that $b_{\rm a}=Va_{\rm a}V^{\dag}$. It remains to find
$V(\Lambda, \bm k)$ given $L_{\textbf{a}}{^\textbf{b}}(\Lambda, \bm k)$. This will be done below.
\subsection{Explicit construction of $V(\Lambda,\bm k)$.}

Consider the generators (\ref{J_1})--(\ref{K_3}).
Construction of $V(\Lambda,\bm k)$ reduces to showing that the unitary operator
\be
V
=\exp\big(i\alpha_{1}L_{1}+i\alpha_{2}L_{2}+i\alpha_{3}L_{3}\big)
\ee
with
\be
\alpha_{1}
&=&\frac{\Theta}{\sin\Theta}
|\phi|
\sin \big(\xi+\Theta\big),
\label{alpha1}\\
\alpha_{2} &=&\frac{\Theta}{\sin\Theta}
|\phi|
\cos\big(\xi+\Theta\big),\label{alpha2}\\
\alpha_{3} &=& 2\Theta,\label{alpha3}
\ee
implies
\be
\left(
\begin{array}{c}
Va_0^{\dag}V^{\dag}\\
Va_1V^{\dag}\\
Va_2V^{\dag}\\
Va_3 V^{\dag}
\end{array}
\right)
&=&
\left(
\begin{array}{cccc}
1+\frac{|\phi|^2}{2} & -|\phi|\cos(\xi+2\Theta) &
|\phi|\sin(\xi+2\Theta) &
-\frac{|\phi|^2}{2}\\
-|\phi|\cos \xi
& \cos 2\Theta & -\sin 2\Theta &
|\phi|\cos \xi\\
|\phi|\sin \xi &
\sin 2\Theta & \cos 2\Theta &
-|\phi|\sin \xi\\
\frac{|\phi|^{2}}{2}
& -|\phi|\cos(\xi+2\Theta)
& |\phi|\sin
(\xi+2\Theta) & 1-\frac{|\phi|^{2}}{2}
\end{array}
\right)
\left(
\begin{array}{c}
a_0^{\dag}\\
a_1\\
a_2\\
a_3
\end{array}
\right).
\ee
The calculation is straightforward. Comparison with (\ref{LLLLL}) ends the proof.
Further simplification is obtained by means of the formulas, valid for (\ref{alpha1})--(\ref{alpha3}),
\be
e^{-i(\alpha_1L_1+\alpha_2L_2+\alpha_3L_3)}
&=&
e^{-i 2\Theta L_3}
e^{-iL_1|\phi|\sin (\xi+2\Theta)}
e^{-iL_2|\phi|\cos (\xi+2\Theta)}
\\
&=&
e^{-iL_1|\phi|\sin \xi}
e^{-iL_2|\phi|\cos \xi}
e^{-i 2\Theta L_3}.
\ee
An analogous reasoning leads to (\ref{U-R}). We leave it as an exercise to the readers.

\end{document}